\documentclass[twocolumn,iop,twocolappendix,numberedappendix,appendixfloats]{openjournal}

\usepackage{amsmath}
\usepackage{amssymb}
\usepackage{graphicx}
\usepackage{booktabs}

\usepackage{siunitx}
\usepackage{enumitem}
\usepackage{acronym}

\usepackage{xcolor}
\definecolor{linkblue}{RGB}{0,0,255}
\usepackage[
  colorlinks=true,
  linkcolor=linkblue,
  citecolor=linkblue,
  urlcolor=linkblue,
  breaklinks=true,
  bookmarksnumbered=true,
  pdftitle={An LSST-DESC Precursor Project: Hyper Suprime-Cam Year 1 3x2pt in Harmonic Space},
  pdfauthor={D. Sanchez-Cid et al.}
]{hyperref}

\usepackage{orcidlink}

\setlist[itemize]{left=2em}
\newacro{DES}[DES]{Dark Energy Survey}
\newacro{HSC}[HSC]{Hyper Suprime-Cam}
\newacro{SDSS}[SDSS]{Sloan Digital Sky Survey}
\newacro{KiDS}[KiDS]{Kilo-Degree Survey}
\newacro{2PCF}[2PCF]{2-point correlation function}
\newacro{LSST}[LSST]{Vera C. Rubin Observatory Legacy Survey of Space and Time}
\newacro{CDM}[CDM]{cold dark matter}
\newacro{CMB}[CMB]{cosmic microwave background}
\newacro{SNe Ia}[SNe Ia]{type Ia supernovae}
\newacro{LSS}[LSS]{large-scale structure}
\newacro{DESC}[DESC]{Dark Energy Science Collaboration}
\newacro{RSD}[RSD]{redshift space distortions}
\newacro{SNR}[SNR]{signal-to-noise ratio}
\newacro{PCA}[PCA]{principal component analysis}
\newacro{tSZ}[tSZ]{thermal Sunyaev--Zel'dovich}
\newacro{MCMC}[MCMC]{Markov Chain Monte Carlo}
\newacro{CCL}[CCL]{Core Cosmology Library}
\newacro{AGN}[AGN]{active galactic nuclei}
\newacro{IA}[IA]{intrinsic alignments}
\newacro{NLA}[NLA]{non-linear alignment}
\newacro{PSF}[PSF]{point spread function}
\newacro{HOD}[HOD]{halo occupation distribution}
\newacro{SSC}[SSC]{super-sampling covariance}
\newacro{cNG}[cNG]{connected non-Gaussian}
\newacro{SP}[SP]{survey property}

\submitted{Version \today}
\journalinfo{Version \today}

\makeatletter
\def\@listand{}
\makeatother

\makeatletter
\def\frontmatter@above@affilgroup{\vspace*{-0.2in}}
\long\def\@makecaption#1#2{%
 \noindent\begin{minipage}{0.9999\linewidth}%
   \if\csname ftype@\@captype\endcsname 2
     \vskip 2ex\noindent\@table@type@size{#1.}\ #2\par\medskip
   \else
     \vspace*{\abovecaptionskip}\noindent\footnotesize #1 #2\par\vskip\belowcaptionskip
   \fi
 \end{minipage}\par
}
\makeatother

\makeatletter

\def\l@f@section{%
  \addpenalty{\@secpenalty}%
  \addvspace{0.8em plus\p@}%
  \bfseries
}%
\def\print@toc#1{%
 \begingroup
  \par\addvspace{0.2\baselineskip}%
  {\noindent\normalsize\csname #1name\endcsname\par}%
  \nobreak\vspace{0.15\baselineskip}%
  \let\appendix\appendix@toc
  \@starttoc{#1}%
 \endgroup
}%
\makeatother

\shorttitle{LSST Precursor: HSC $3\times2$pt in harmonic space}
\shortauthors{Sanchez-Cid et al.}

\begin{document}

\title[LSST Precursor: HSC $3 \times 2$pt in harmonic space]{An LSST-DESC Precursor Project: Hyper Suprime-Cam Year 1 3 $\times$ 2pt in Harmonic Space}

\author{
D. Sanchez-Cid\orcidlink{0000-0003-3054-7907}\altaffilmark{1,2,$\star$},
J. Sanchez\orcidlink{0000-0003-3136-9532}\altaffilmark{3},
I. Sevilla-Noarbe\orcidlink{0000-0002-1831-1953}\altaffilmark{1},
D. Alonso\orcidlink{0000-0002-4598-9719}\altaffilmark{4},
F. Andrade-Oliveira\orcidlink{0000-0003-0171-6900}\altaffilmark{2},
H. Awan\orcidlink{0000-0003-2296-7717}\altaffilmark{5,6},
C. Chang\orcidlink{0000-0002-7887-0896}\altaffilmark{7,8,9},
J. Ellison\orcidlink{0000-0001-8692-9504}\altaffilmark{10},
C. Garc\'ia-Garc\'ia\orcidlink{0000-0001-6394-7494}\altaffilmark{1,4,11,12},
M. Ishak\orcidlink{0000-0002-6024-466X}\altaffilmark{13},
E. Longley-Phillips\altaffilmark{14},
R. Mandelbaum\orcidlink{0000-0003-2271-1527}\altaffilmark{15},
A. Nicola\orcidlink{0000-0003-2792-6252}\altaffilmark{16},
E. Pedersen\orcidlink{0000-0002-8883-2172}\altaffilmark{17},
J. Prat\orcidlink{0000-0002-5933-5150}\altaffilmark{18,19},
E. Rykoff\orcidlink{0000-0001-9376-3135}\altaffilmark{5,6},
E. Sanchez\orcidlink{0000-0002-9646-8198}\altaffilmark{1},
N. \v{S}ar\v{c}evi\'c\orcidlink{0000-0001-7301-6415}\altaffilmark{14},
M. Soares-Santos\orcidlink{0000-0001-6082-8529}\altaffilmark{2},
M. Yamamoto\orcidlink{0000-0003-1585-997X}\altaffilmark{20},
J. Zuntz\orcidlink{0000-0001-9789-9646}\altaffilmark{21},
and the LSST Dark Energy Science Collaboration
}
\affil{Author affiliations may be found before the references.}

\altaffiltext{$\star$}{E-mail: david.sanchezcid@physik.uzh.ch}

\begin{abstract}
We present the first fully photometric joint analysis of weak gravitational lensing and galaxy clustering (3 $\times$ 2pt) in harmonic space with LSST-DESC analysis pipelines applied to Hyper Suprime-Cam Year~1 (HSC~Y1) data. The HSC~Y1 dataset, with imaging depth and galaxy number density similar to those expected for the first year of LSST observations, is an ideal testbed for validating DESC measurement and inference tools. We measure the full set of angular power spectra --- cosmic shear, galaxy clustering, and galaxy--galaxy lensing --- perform null tests, and correct for the impact of observing conditions on both the galaxy density and shear fields via mode deprojection. We validate our likelihood pipeline by reproducing the official HSC~Y1 cosmic shear cosmological constraints with the DESC inference code. Jointly analysing all three two-point functions, we constrain $\Lambda$CDM and $w$CDM cosmologies, reporting results for cosmic shear, the combination of galaxy clustering and galaxy--galaxy lensing (2 $\times$ 2pt), and the full 3 $\times$ 2pt. From the joint 3 $\times$ 2pt $\Lambda$CDM analysis we find $\Omega_{\rm m} = 0.249^{+0.061}_{-0.051}$, $\sigma_8 = 0.904^{+0.098}_{-0.091}$, and $S_8 \equiv \sigma_8\sqrt{\Omega_{\rm m}/0.3} = 0.821^{+0.018}_{-0.021}$. This analysis validates the complete DESC 3 $\times$ 2pt infrastructure on survey data, providing the foundation for the forthcoming LSST Year~1 cosmological analysis.
\end{abstract}


\makeatletter
\maketitle
\setcounter{footnote}{\thefront@matter@foot@note}
\let\footnotetext=\old@foot@note@text
\let\footnotemark=\old@foot@note@mark
\@firstsectionfalse
\makeatother


\section{Introduction}
\label{sec:introduction}

The nature of the accelerated expansion of the Universe at late times and the growth of \ac{LSS} remain among the most fundamental open questions in modern cosmology. The standard cosmological model, $\Lambda$CDM, addresses both  through two key ingredients: a cosmological constant $\Lambda$ as a source of accelerated expansion, and \ac{CDM} as the seed of \ac{LSS}. This framework is supported by foundational observations of the \ac{CMB}~\citep{aghanim2020planck} and of \ac{SNe Ia}~\citep{riess1998observational, perlmutter1999measurements, schmidt1998high, betoule2014improved}. Complementary probes tracing the growth of structure and the geometry of the Universe have since provided increasingly precise tests of this model.

In the last two decades, the joint analysis of weak gravitational lensing and galaxy clustering (the so-called $3 \times 2$pt) has emerged as one of the most powerful tools for constraining cosmological models, driven by Stage~III galaxy surveys \citep{albrecht2006report}. The \ac{DES} Collaboration has released its legacy cosmology results from cosmic shear \citep{DES:2026mkc} and $3 \times 2$pt \citep{DES:2026fyc} analyses in configuration space, complemented by harmonic-space analyses of earlier datasets \citep{Andrade_Oliveira_2021, Doux_2022, faga2024darkenergysurveyyear}. The \ac{HSC} Collaboration has constrained cosmology with cosmic shear \citep{li2023hyper, dalal2023hyper} and with a $3 \times 2$pt analysis using spectroscopic \ac{SDSS} DR11 lens galaxies \citep{Sugiyama_2023, miyatake2023hyper}, building on the HSC Year~3 data release \citep{Li_2022}. The \ac{KiDS} has contributed cosmic shear legacy results \citep{Wright:2025xka} and a $3 \times 2$pt analysis with \ac{SDSS} lenses \citep{Heymans:2020gsg}. Together, these Stage~III surveys establish the methodology and datasets that underpin the design of Stage~IV projects: the \ac{LSST} \citep{lsst2012large}, the \textit{Euclid} mission \citep{Euclid:2024lrr}, and the \textit{Nancy Grace Roman Space Telescope} \citep{Spergel:2015sza,Akeson:2019biv}, among others.

The coming generation of cosmic surveys will require end-to-end analysis pipelines capable of condensing galaxy catalogs into summary statistics at unprecedented scale, alongside flexible statistical inference codes able to accommodate increasingly complex theoretical models. In preparation for \ac{LSST}, we present a precursor $3 \times 2$pt analysis of the \ac{HSC} Y1 dataset with \ac{LSST}-\ac{DESC} analysis software, extending the cosmic shear re-analysis of \citet{Longley_2023} to the full joint weak lensing and galaxy clustering analysis. The $3 \times 2$pt method is central to the \ac{DESC} cosmology program, targeting constraints on $\Omega_{\rm m}$ and $S_8 \equiv \sigma_8 \sqrt{\Omega_{\rm m}/0.3}$.

In this work, we measure the angular power spectra of cosmic shear, galaxy--galaxy lensing, and galaxy clustering from the \ac{HSC} Y1 galaxy catalog using \textsc{TXPipe} \citep{prat2022catalog}, validating the measurements against existing results and presenting a suite of null tests. We correct for spurious contributions from observing conditions to the galaxy density and shear fields via mode deprojection and compute an analytical covariance matrix with \textsc{TJPCov}. Cosmological constraints are derived with \textsc{Firecrown} \textit{(in prep.)} for the cosmic shear, $2 \times 2$pt, and $3 \times 2$pt analyses; we validate the pipeline by reproducing the cosmic shear constraints from the official \ac{HSC} Y1 analysis \citep{hikage2019}.

The paper is organized as follows. Section~\ref{section:data} describes the \ac{HSC} Y1 data and the lens and source galaxy sample selection. Section~\ref{sec:theory} presents the theoretical framework for the $3 \times 2$pt analysis. Section~\ref{section:measurements} presents the angular power spectra measurements, null tests and mitigation of observing condition systematics. Section~\ref{sec:modeling} describes the likelihood pipeline, modeling choices and scale cuts. Section~\ref{section:analysis} introduces the parameter inference framework. Cosmological constraints on $\Lambda$CDM and $w$CDM, including extensions to massive neutrinos, and robustness tests are presented in Section~\ref{section:results}. Finally, Section~\ref{section:conclusions} summarizes our findings and outlines implications for the first \ac{LSST}-\ac{DESC} data analysis.

\section{Data}
\label{section:data}

\ac{HSC} is a multiband imaging survey that acquired data from 2014 to 2019, covering a final area of \num{1400}~deg$^2$ of the northern sky in five broad photometric bands (\textit{grizy}), using the Hyper Suprime-Cam mounted on the Subaru Telescope. The Wide layer offers a $5\sigma$ depth for point sources of $i = 26.2$ and number density of $22.9$ galaxies per arcmin$^2$, making it well suited for weak lensing measurements. The first public data release \citep{aihara2018first} was processed using a prototype version of the Rubin Science Pipelines \citep{juric2015lsst, bosch2018hsc}, and represents one of the deepest wide-field imaging catalogs available, with source number densities that foreshadow the statistical challenges expected with \ac{LSST} --- making it a natural precursor dataset for \ac{LSST}-\ac{DESC}.

In this work, we analyse the first-year shear catalog \citep{mandelbaum2017hsc}, comprising positions and shapes of approximately 11 million galaxies over 136.9~deg$^2$ across five non-contiguous fields: \textsc{Gama09h}, \textsc{Gama15h}, \textsc{Vvds}, \textsc{Wide12h}, and \textsc{Xmm}; we exclude \textsc{Hectomap} due to its small area, following \citet{nicola2020tomographic}. Data reduction and catalog cuts were performed independently of the official HSC analysis.

\subsection{Source galaxies}
\label{subsec:source-sample}

The catalog applies selection criteria optimized for ellipticity measurement and shear calibration uniformity; galaxy shapes are measured in the $i$-band using the re-Gaussianization method, while photometric redshifts are estimated from all five \textit{grizy} bands (Section~\ref{subsec:photo-z}). See Section~2.2 of~\citet{mandelbaum2017hsc} for details.

The sample is limited to $i < 24.5$, a magnitude threshold consistent with what can plausibly be applied to the first one to two years of LSST data \citep{alonso2018lsst}, yielding a number density of $17.9$ galaxies per arcmin$^2$. Tomographic redshift bins are defined following \citet{hikage2019} using the \textsc{Ephor\_ab} photo-$z$ code which is described below, spanning $0.3 < z < 1.5$ with source number densities of $\{5.6,\, 5.6,\, 4.2,\, 2.5\}$ galaxies per arcmin$^2$, totaling \num{8813360} objects over 136.9~deg$^2$. A breakdown by field and tomographic bin is given in Appendix~\ref{app:galaxy-numbers}.

\subsection{Lens galaxies}
\label{subsec:lens-sample}

The lens galaxy sample is selected following \citet{nicola2020tomographic}, using the \textsc{DR1} version of the dataset covering 93~deg$^2$. Selection cuts (summarized in Table~1 of \citealt{nicola2020tomographic}) ensure accurately measured fluxes across available bands and exclude artifacts, stars, and spurious detections. A magnitude threshold of $i_{\rm corr} < 24.5$ defines a homogeneous, complete sample of objects detected in at least two bands. The resulting magnitude-limited sample comprises \num{8235621} objects in four tomographic redshift bins spanning $0.15 < z < 1.5$, with a number density of $24.6$ galaxies per arcmin$^2$ and per-bin densities of $\{6.0,\, 6.3,\, 6.0,\, 6.3\}$ galaxies per arcmin$^2$. The survey geometry mask is constructed following \citet{nicola2020tomographic}, combining the HSC DR1 bright-object mask --- which excludes regions around bright stars using magnitude-dependent exclusion radii --- with a depth cut removing pixels with $10\sigma$ limiting depth below $i < 24.5$.

\subsection{Redshift distributions}
\label{subsec:photo-z}

\begin{figure}[t]
    \centering
    \includegraphics[width=0.85\linewidth]{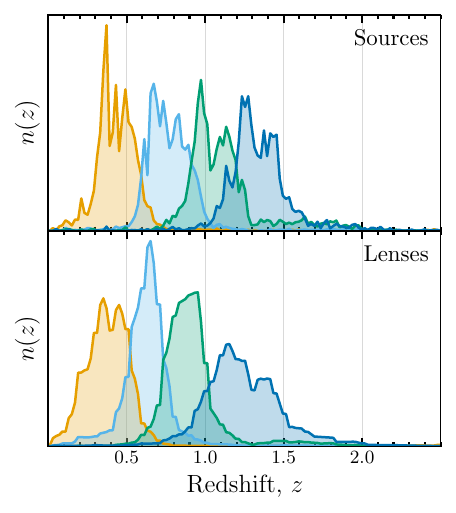}
    \caption{HSC Y1 normalized redshift distributions estimated with \textsc{Ephor\_ab}. Source galaxies are divided into four redshift bins with edges $[0.3, 0.6, 0.9, 1.2, 1.5]$, and lens galaxies into four bins with edges $[0.15, 0.5, 0.75, 1.0, 1.5]$.}
    \label{fig:dndz}
\end{figure}

We estimate redshift distributions following the official \ac{HSC} Y1 methodology \citep{tanaka2018photometric}, using the machine learning-based photo-$z$ code \textsc{Ephor\_ab}. The network takes as input de Vaucouleurs and exponential model fluxes from each band, supplemented by PSF-matched aperture photometry. The resulting redshift distributions for the source and lens samples are shown in Figure~\ref{fig:dndz}. We additionally obtain estimates from \textsc{Franken-z} and \textsc{Nnz} to assess the robustness of our cosmological constraints to the choice of photo-$z$ code (Section~\ref{subsec:results-robustness}).

\section{Theory}
\label{sec:theory}

The $3 \times 2\mathrm{pt}$ analysis combines auto- and 
cross-correlations of the galaxy density contrast $\delta^i_g$ and the shear field $\gamma^j_\alpha$ ($\alpha \in \{1,2\}$ labelling the two spin-2 shear components) across tomographic redshift bins $i$ and $j$. The normalized redshift distribution of galaxy samples in bin $i$ is $n^i_g(z)$, where $g = l, s$ denotes the lens and source samples, respectively.

The observed galaxy density contrast along direction $\hat{n}$ receives three contributions,
\begin{equation}
    \delta^i_g(\hat{n}) = \delta^i_{g,\mathrm{D}}(\hat{n}) 
    + \delta^i_{g,\mathrm{RSD}}(\hat{n}) 
    + \delta^i_{g,\mu}(\hat{n}),
\end{equation}
where 
\begin{equation}
    \delta^i_{g,\mathrm{D}}(\hat{n}) = 
    \int d\chi\, W^i_{\delta,g}(\chi)\, \delta_g(\hat{n}\chi, \chi)
\end{equation}
is the projected galaxy density contrast with tomographic clustering kernel
\begin{equation}
    W^i_{\delta,g}(\chi) = n^i_g(z)\frac{dz}{d\chi},
\end{equation}
$\chi$ is the comoving distance, $\delta^i_{g,\mathrm{RSD}}$ is the \ac{RSD} term, and $\delta^i_{g,\mu}$ accounts for lens magnification \citep{Joachimi:2009ez, DES:2016lyd} via variations in the projected number density and magnification effects on galaxy flux and size. The magnification kernel is
\begin{equation}
    W^i_\mu(\chi) = \frac{3\Omega_{\rm m}H_0^2}{2c^2}
    \frac{\chi}{a(\chi)}
    \int_\chi^\infty \mathrm{d}\chi'\, n^i_{g}(\chi')\,
    \bigl(5s(\chi', m_{\rm lim})-2\bigr)
    \frac{\chi'-\chi}{\chi'},
    \label{eq:magnif-kernel}
\end{equation}
where $s$ is the slope of the cumulative apparent magnitude distribution,
\begin{equation}
    s(\chi, m_{\rm lim}) \equiv
    \left.\frac{\partial \log_{10} N(<m,\chi)}{\partial m}
    \right|_{m = m_{\rm lim}}.
    \label{eq:magnif-slope}
\end{equation}

The baseline analysis includes the $\delta_{g,\mathrm{D}}$ and $\delta_{g,\mathrm{RSD}}$ terms; the magnification contribution is explored as an extension in Section~\ref{subsec:results-extended}.

The observed shear field $\gamma^j_\alpha$ has two main contributions: gravitational shear and intrinsic ellipticity. The latter comprises a spatially coherent component from \ac{IA} and shape noise $\epsilon_0$,
\begin{equation}
    \gamma^j_\alpha(\hat{n}) = \gamma^j_{\alpha,\mathrm{G}}(\hat{n}) 
    + \gamma^j_{\alpha,\mathrm{IA}}(\hat{n}) 
    + \epsilon^j_{\alpha,0}(\hat{n}).
\end{equation}
The shape noise $\epsilon_0$ contributes to the covariance matrix but not to the two-point signal. Cosmic shear is decomposed into E- and B-modes as
\begin{align}
    C^{ij}_{EE}(\ell) &= C^{ij}_{\kappa_g \kappa_g}(\ell) 
    + C^{ij}_{\kappa_g I_E}(\ell) 
    + C^{ji}_{\kappa_g I_E}(\ell) 
    + C^{ij}_{I_E I_E}(\ell), \\
    C^{ij}_{BB}(\ell) &= C^{ij}_{I_B I_B}(\ell),
\end{align}
where $C^{ij}_{BB}$ should be consistent with zero in the absence of observational systematics, and $I_{E/B}$ denotes the E-/B-modes of the \ac{IA}. The galaxy--galaxy lensing power spectrum is

\begin{align}
    C^{ij}_{\delta_g E}(\ell) &= 
      C^{ij}_{\delta_{g,\mathrm{D}}\kappa_g}(\ell) 
    + C^{ij}_{\delta_{g,\mathrm{D}}I_E}(\ell) \nonumber \\
    &+ C^{ij}_{\delta_{g,\mu}\kappa}(\ell) 
    + C^{ij}_{\delta_{g,\mu}I_E}(\ell) \nonumber \\
    &+ C^{ij}_{\delta_{g,\mathrm{RSD}}\kappa_g}(\ell) 
    + C^{ij}_{\delta_{g,\mathrm{RSD}}I_E}(\ell).
\end{align}

The angular cross-power spectrum between two fields $\mathcal{A}$ and $\mathcal{B}$ is estimated under the Limber approximation as

\begin{equation}
    C^{ij}_{\mathcal{A}\mathcal{B}}(\ell) = \int \frac{d\chi}{\chi^2}\,
    W^i_{\mathcal{A}}(\chi)\,W^j_{\mathcal{B}}(\chi)\,
    P_{\mathcal{A}\mathcal{B}}\!\left(k=\frac{\ell+1/2}{\chi},\,z(\chi)\right),
\end{equation}

where $P_{\mathcal{A}\mathcal{B}}(k,z)$ is the 3D power spectrum and $W_A$, $W_B$ are the corresponding projection kernels. For galaxy clustering, $C_{\delta_g\delta_g}$ is computed using the full non-Limber expressions \citep{krause2021dark, Fang_2020}.

\section{Angular power spectra measurement}
\label{section:measurements}

We measure cosmic shear, galaxy clustering, and galaxy--galaxy lensing angular power spectra using \textsc{TXPipe}~\citep{prat2022catalog}, the DESC end-to-end pipeline for two-point statistics in real and harmonic space. \textsc{TXPipe} ingests galaxy shape and lens catalogs, splits galaxies into tomographic redshift bins, estimates noise maps, masks spurious regions, and interfaces with \textsc{TJPCov} for the analytical covariance estimate (Section~\ref{subsec:covariance}). 

Angular power spectra are measured accounting for the partial-sky coverage of the \ac{HSC} footprint using the pseudo-$C_\ell$ methodology \citep{Alonso_2019, Hivon_2002}, which handles mode-coupling effects arising from incomplete sky coverage. Masks and their associated mode-coupling matrices are  computed separately for the lens and source samples rather than as a  single joint mask, and are common to all tomographic bins within each sample. No apodization is applied; the mask-induced mode coupling is accounted for analytically by deconvolving the mode-coupling matrix within the pseudo-$C_\ell$ framework. We adopt multipole bin edges $[100, 200, 300, 400, 600, 800, 1000, 1400, 1800, 2200]$, following \citet{hikage2019} and \citet{nicola2020tomographic} to facilitate comparison. The full $3 \times 2\mathrm{pt}$ data vector comprises 4 galaxy clustering auto-spectra --- cross-correlations between different lens bins are excluded, as their signal is dominated by photometric redshift leakage between bins and receives additional magnification contributions that complicate the modeling, with modest gain in constraining power (see Section VI A of~\cite{Porredon_2022}) --- 10 cosmic shear auto- and cross-spectra, and 16 galaxy--galaxy lensing cross-spectra, totaling 270 data points before scale cuts (Section~\ref{subsec:cuts}). The \ac{SNR} is computed as

\begin{equation}
    \mathrm{SNR} \equiv \sqrt{\hat{D}^\intercal\, 
    \mathrm{Cov}^{-1}\, \hat{D}},
\end{equation}

where $\mathrm{Cov}$ is the analytical covariance matrix from \textsc{TJPCov} (Section~\ref{subsec:covariance}) and

\begin{equation}
    \hat{D} \equiv \left\{\hat{C}_{EE}(\ell),\, \hat{C}_{\delta_g E}(\ell),\, \hat{C}_{\delta_g \delta_g}(\ell)\right\}
\end{equation}

is the concatenated data vector. After applying scale cuts (Section~\ref{subsec:cuts}), the data vector reduces to 115 points --- 60 from cosmic shear, 44 from galaxy--galaxy lensing, and 11 from galaxy clustering --- for a total $\mathrm{SNR} = 44$.

\subsection{$C_\ell$ comparison with the literature and null tests}
\label{subsection:meas-consistency}

\begin{figure*}
    \centering
    \includegraphics[width=0.72\textwidth]{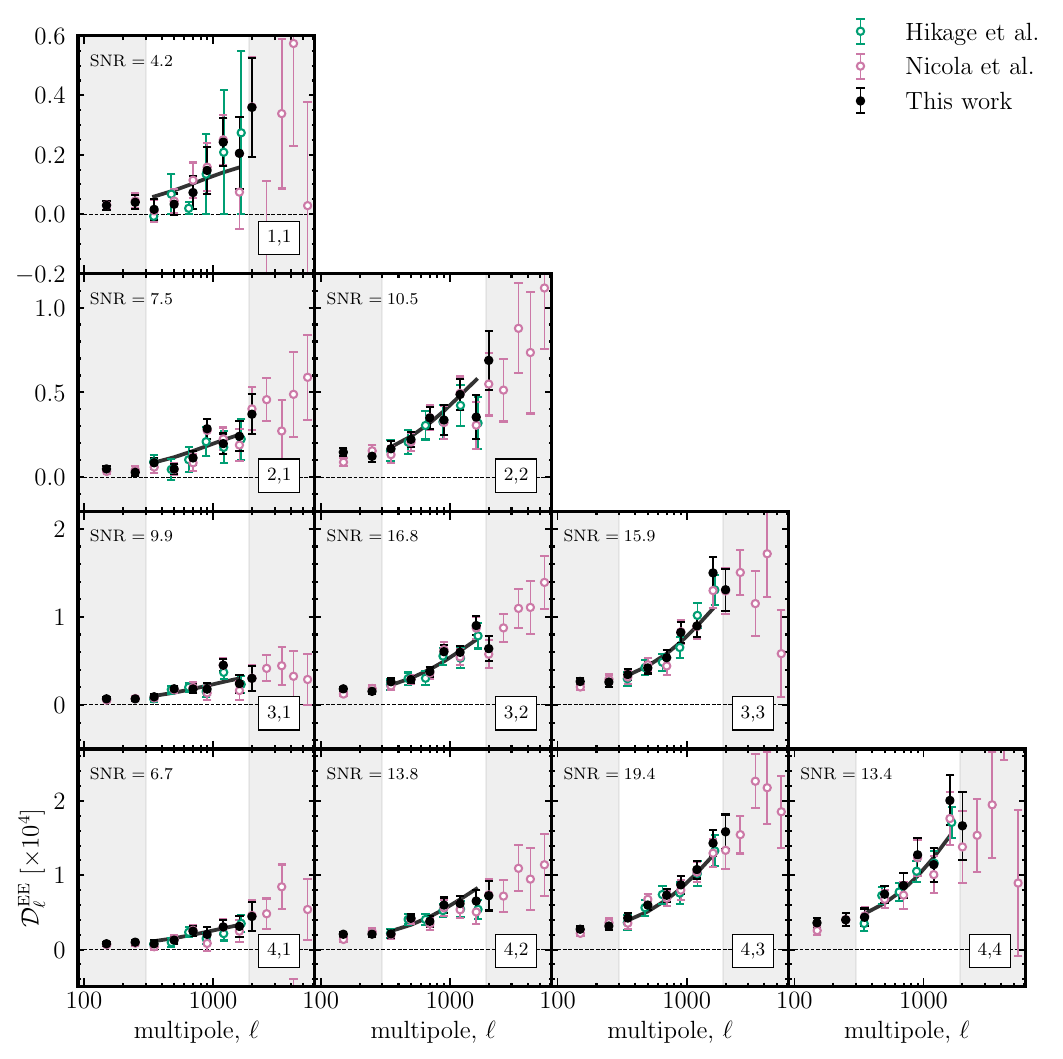}
    \includegraphics[width=0.75\linewidth]{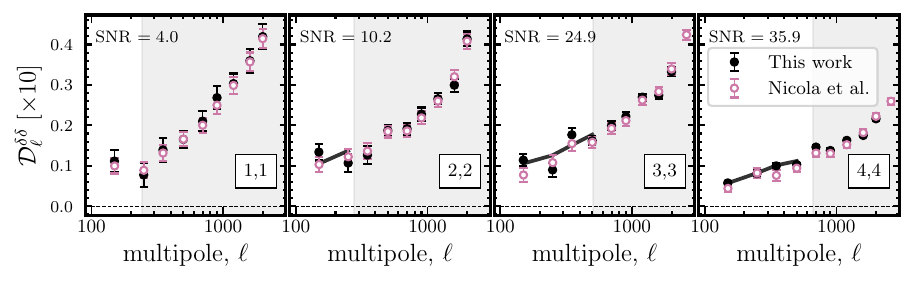}
    \caption{Angular power spectra measured in this work (black filled circles) compared to results from the literature. \textit{Top:} Cosmic shear spectra scaled as $\mathcal{D}_\ell = \ell(\ell+1)C_\ell/2\pi$, compared to \citet{hikage2019} (open green circles) and \citet{nicola2021cosmic} (open pink circles), both using the pseudo-$C_\ell$ method. Numeric pairs label the tomographic bin combinations. \textit{Bottom:} Galaxy clustering auto-spectra, compared to \citet{nicola2020tomographic} (open pink circles). In both panels, the solid black line shows the theoretical prediction at the $3\times2$pt best-fit model, and gray shaded regions indicate scales excluded from the analysis: $\ell < 300$ and $\ell > 1900$ for cosmic shear, and $k > 0.15\,\mathrm{Mpc}^{-1}$ for galaxy clustering. In the $(1,1)$ panel a single multipole bin remains after scale cuts; the theory line is omitted to avoid implying the joint best-fit cosmology was constrained from this panel alone. In the $(3,3)$ and $(4,4)$ panels, filled black circles (this work) may overlap with the open pink circles \citep{nicola2020tomographic} at the highest retained multipole bin.}
    \label{fig:shear-cross-literature}
    \label{fig:meas-clustering-dr1}
\end{figure*}

We validate our angular power spectra measurements against independent pipelines. Cosmic shear auto-correlations are compared to \citet{hikage2019} and \citet{nicola2021cosmic}, both using the pseudo-$C_\ell$ method, finding good agreement as shown in Figure~\ref{fig:shear-cross-literature}. Galaxy clustering spectra are compared to \citet{nicola2020tomographic}, also finding good agreement (Figure~\ref{fig:meas-clustering-dr1}). For both comparisons the galaxy samples are identical, but differences are expected due to the distinct pixelization schemes and pseudo-$C_\ell$ implementations: while \citet{hikage2019} and \citet{nicola2020tomographic} adopt a flat-sky approach, defining rectangular regions around each HSC field and applying fast Fourier transforms independently per field, we use full-sky \textsc{HEALPix} maps \citep{Gorski_2005} with a curved-sky pseudo-$C_\ell$ estimator applied jointly across all fields.

\begin{figure*}
    \centering
    \includegraphics[width=0.8\textwidth]{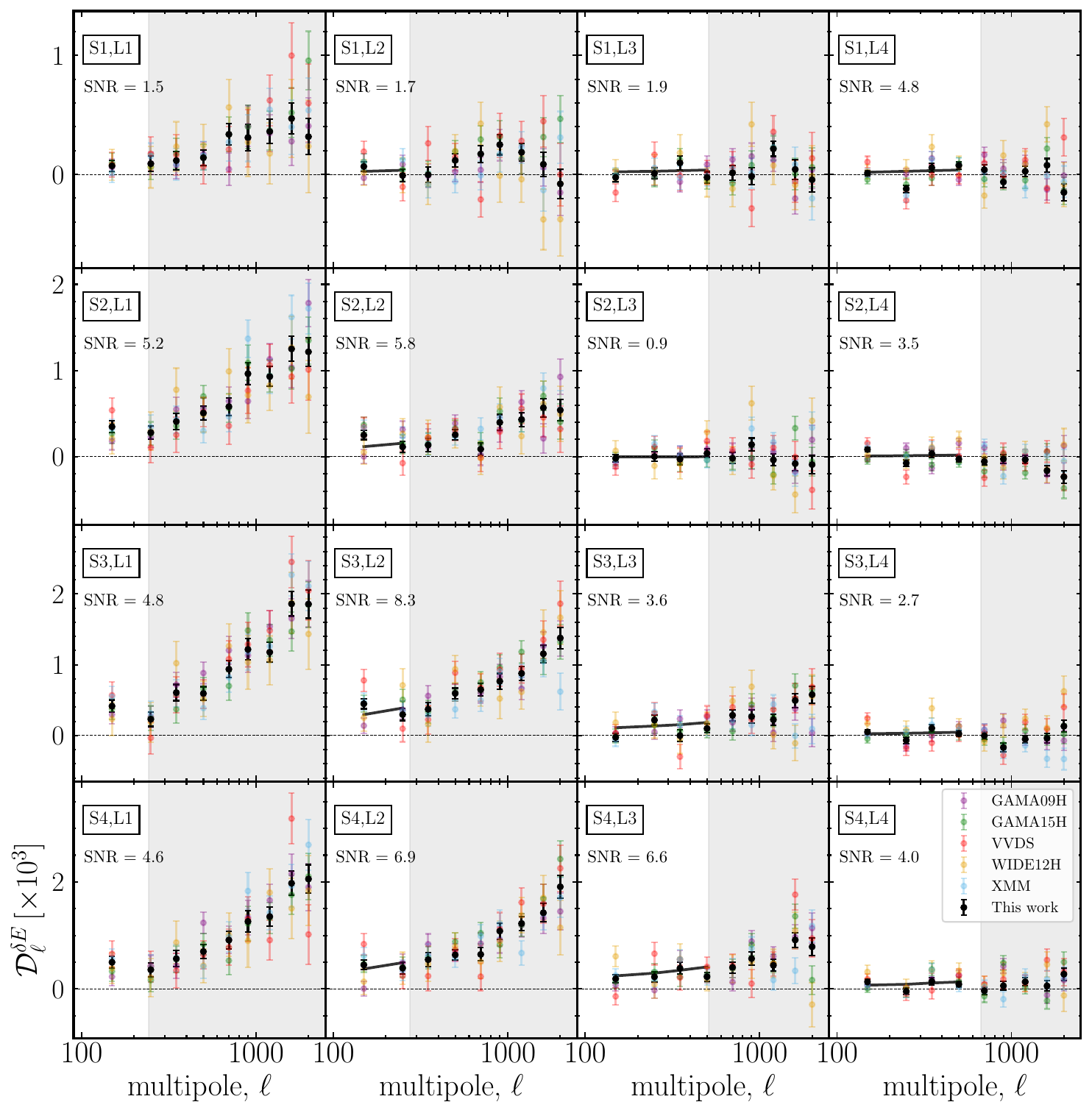}
    \caption{Galaxy--galaxy lensing angular power spectra $\mathcal{D}_\ell^{\delta E} = \ell(\ell+1)C_\ell^{\delta E}/2\pi$ measured with \textsc{TXPipe}. Black filled circles show the combined full-footprint measurement; colored markers show the five individual fields ({\sc Gama09h}, {\sc Gama15h}, {\sc Vvds}, {\sc Wide12h}, {\sc Xmm}). Labels of the form Si,Lj identify source redshift bin $i$ and lens redshift bin $j$. The solid black line shows the theoretical prediction at the $3\times2$pt best-fit model. Gray shaded regions indicate scales excluded from the analysis at $k > 0.15\,\mathrm{Mpc}^{-1}$.}
    \label{fig:gglensing-measurement}
\end{figure*}

\begin{figure*}
    \centering
    \includegraphics[width=0.49\textwidth]{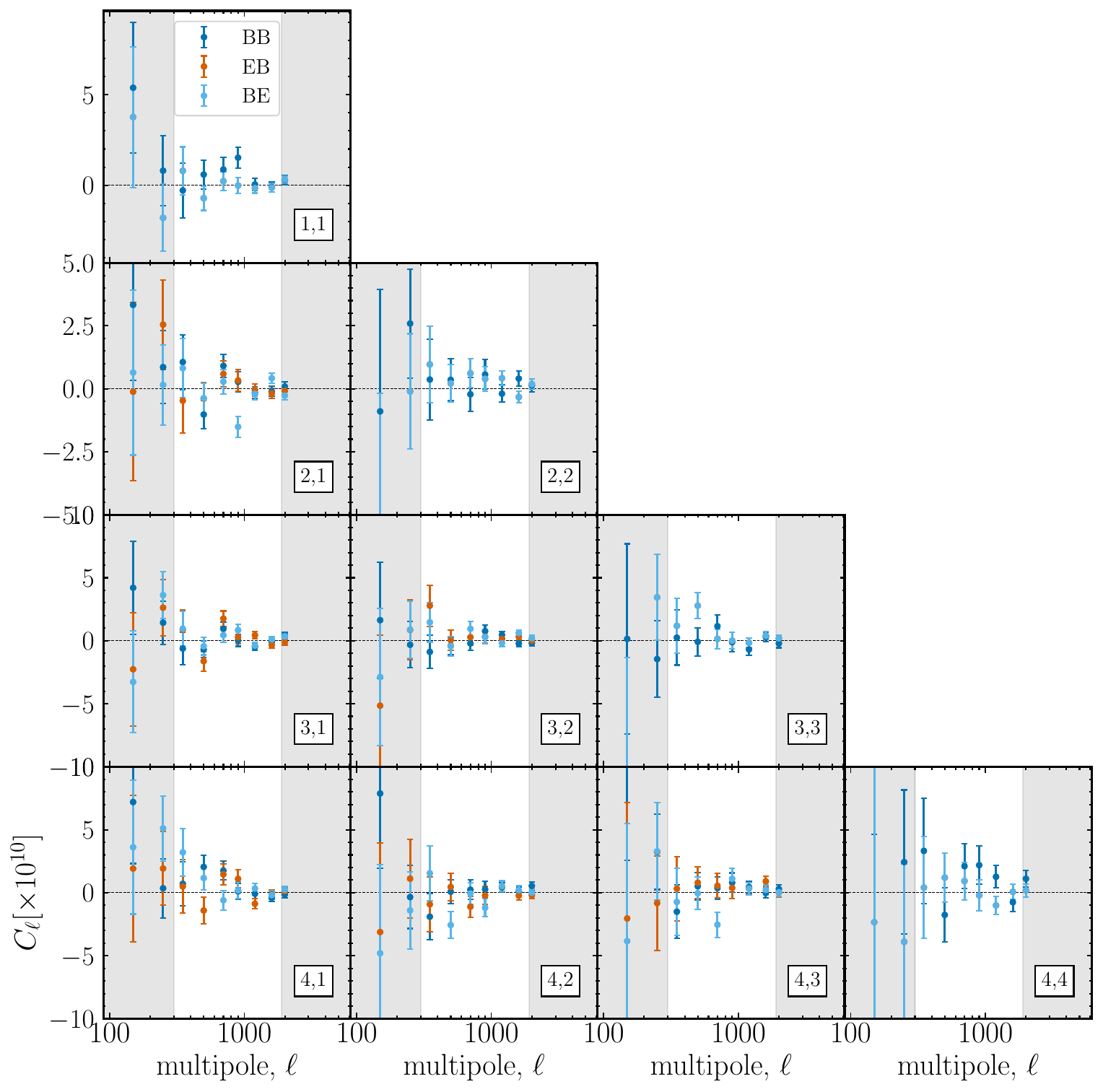}
    \includegraphics[width=0.46\textwidth]{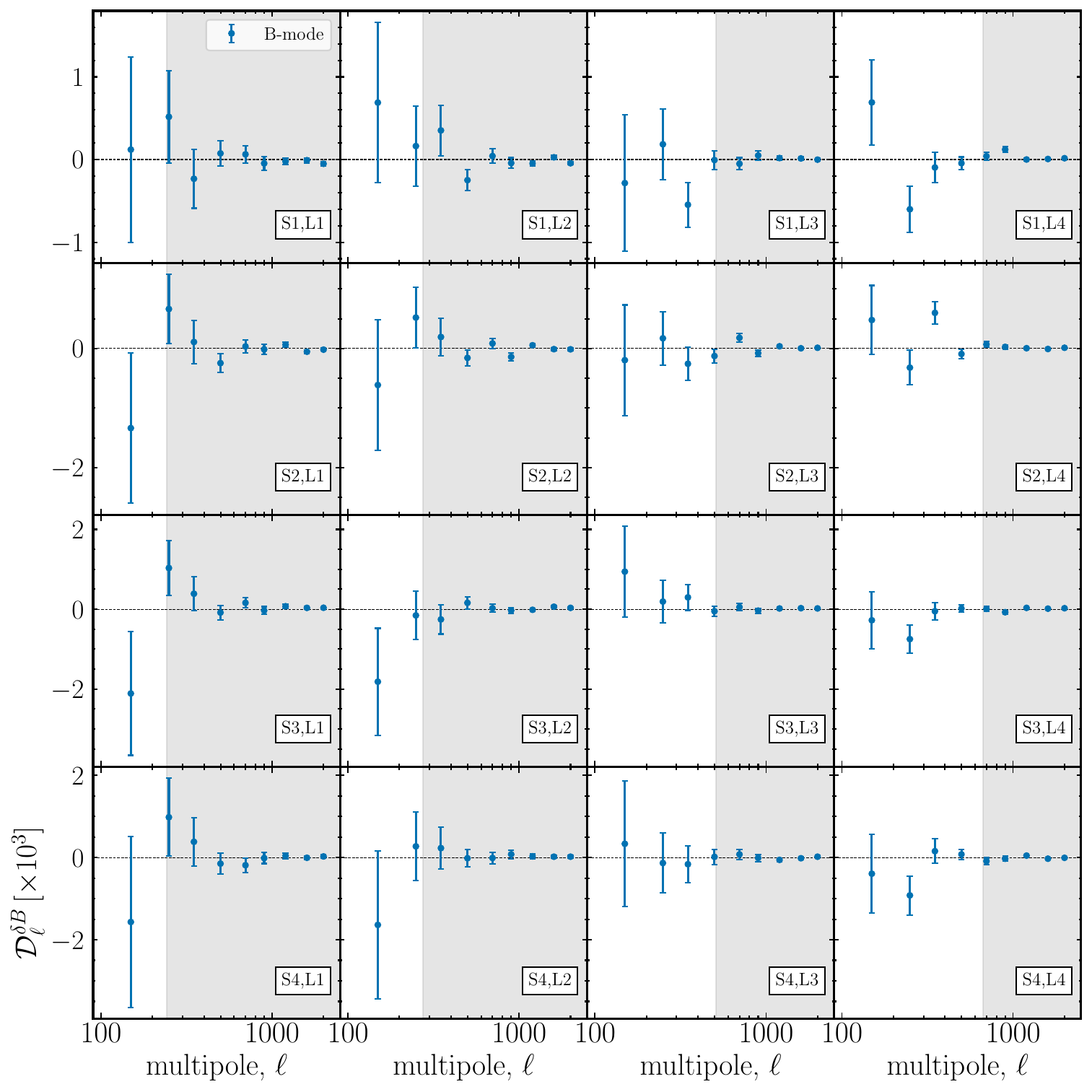}
    \caption{B-mode null tests for cosmic shear and galaxy--galaxy lensing. \textit{Left:} BB- (dark blue), EB- (orange), and BE-mode (light blue) power spectra $C_\ell$ from the cosmic shear measurement, for each tomographic bin-pair $(i,j)$. The BB and BE modes are consistent with zero within the analyzed scales; the EB mode shows a marginally significant deviation. \textit{Right:} B-mode power spectra $\mathcal{D}_\ell^{\delta B}$ from the galaxy--galaxy lensing measurement, for each source--lens bin combination Si,Lj. The signal is consistent with zero (see text for $\chi^2$ values).}
    \label{fig:null-test}
\end{figure*}

We also measure the galaxy--galaxy lensing angular power spectra (Figure~\ref{fig:gglensing-measurement}), showing the full-footprint signal alongside the five individual field measurements. The B-mode null test yields $\chi^2/\nu = 162/144$ (p = 0.14), showing no statistically significant evidence for B-mode contamination (Figure~\ref{fig:null-test}).

We check for B-mode contamination in the cosmic shear signal, which would indicate the presence of non-gravitational systematics. The BB, EB, and BE modes shown in Figure~\ref{fig:null-test} are evaluated after applying scale cuts: $\chi^2/\nu = 68.9/60$ (p = 0.20) for BB, $\chi^2/\nu = 65.5/60$ (p = 0.29) for BE, and $\chi^2/\nu = 88.3/60$ (p = 0.01) for EB. The BB and BE modes are consistent with zero; the EB result is marginally significant but within the range expected from statistical fluctuations given the number of null tests performed.

\subsection{Deprojection of observing conditions}
\label{subsec:meas-deprojection}

Imaging surveys are susceptible to observational conditions such as seeing and airmass, which can introduce spurious contributions into the measured power spectra. We isolate the cosmological signal from these contaminants using \ac{SP} maps generated with \textsc{Decasu}. We have 12 scalar and tensor SP maps across the 5 photometric bands (\textit{grizy}); a full list is given in Table~\ref{tab:sp-maps}.

\begin{table}
    \centering
    \small
    \begin{tabular}{ll}
    \toprule
    Field & Observing condition \\
    \midrule
    Galaxy field & \parbox[t]{0.62\columnwidth}{\raggedright\texttt{airmass\_wmean},
                    \texttt{dcr\_ddec\_wmean},
                    \texttt{dcr\_dra\_wmean},
                    \texttt{exptime\_sum},
                    \texttt{maglim\_wmean},
                    \texttt{nexp\_sum},
                    \texttt{seeing\_wmean},
                    \texttt{sigma\_sky\_wmean},
                    \texttt{skylevel\_wmean},
                    \texttt{weight\_sum}} \\
    \midrule
    Shear field   & \parbox[t]{0.62\columnwidth}{\raggedright\texttt{dcr\_e1}, \texttt{dcr\_e2}} \\
    \bottomrule
    \end{tabular}
    \caption{\ac{SP} maps generated with \textsc{Decasu} \citep[following][]{DES:2015vnr, nicola2020tomographic} used for mode deprojection. Each map is estimated across the 5 photometric bands (\textit{grizy}) and characterized by its mean or summed value.}
    \label{tab:sp-maps}
\end{table}

Mode deprojection \citep{elsner2016unbiased, Alonso_2019} subtracts from each observed field the component linearly correlated with the \ac{SP} maps, and accounts for this operation in the mode-coupling matrix to produce unbiased pseudo-$C_\ell$ estimates. We implement this through \textsc{TXPipe} wrapping \textsc{NaMaster} \citep{Alonso_2019}. \ac{SP} maps are generated with \textsc{Decasu} following the methodology of \citet{DES:2015vnr} and as described in \citet{nicola2020tomographic}; comparative analyses of systematics filtering techniques are discussed in \citet{weaverdyck2021mitigating} and \citet{berlfein2024multiplicative}.

For the shear field, we deproject maps encoding Differential Chromatic Refraction (\texttt{dcr\_e1}, \texttt{dcr\_e2}; Table~\ref{tab:sp-maps}), a spin-2 field describing wavelength-dependent atmospheric refraction that distorts galaxy shape measurements. Estimates are split into two components across the five photometric bands, yielding 10 SP maps for the shear field. For the density field, we deproject 10 scalar observing conditions across the 5 photometric bands, yielding 50 SP maps encoding quantities such as airmass, seeing, and exposure time. Deprojecting the full set of 60 maps is computationally expensive under the curved-sky scenario, so we apply \ac{PCA} to the scalar maps to reduce their dimensionality, leaving the spin-2 maps unchanged. We retain the 22 principal components accounting for 98\% of the variance of the original set, following the choice adopted in~\cite{RodriguezMonroy:2021rzt}.

\begin{figure*}
    \centering
    \includegraphics[width=0.49\linewidth]{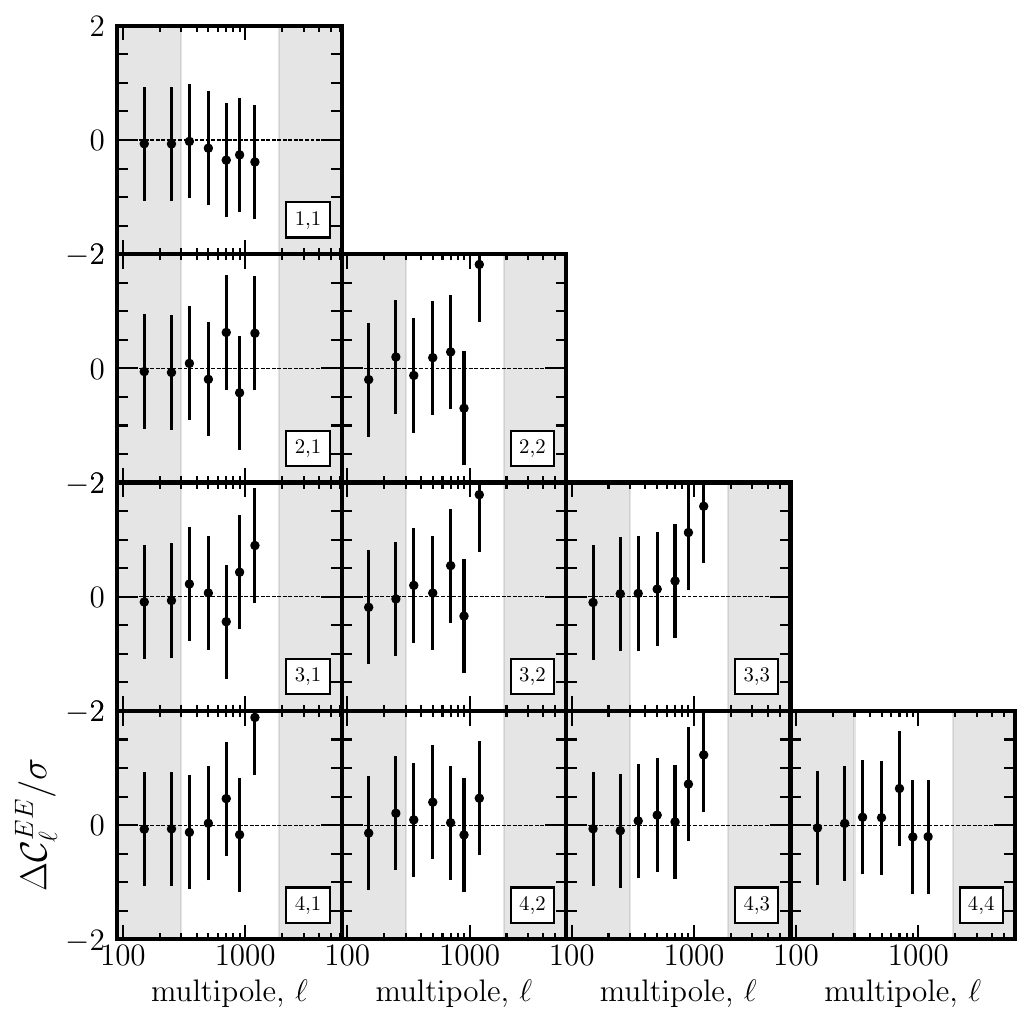}
    \includegraphics[width=0.49\linewidth]{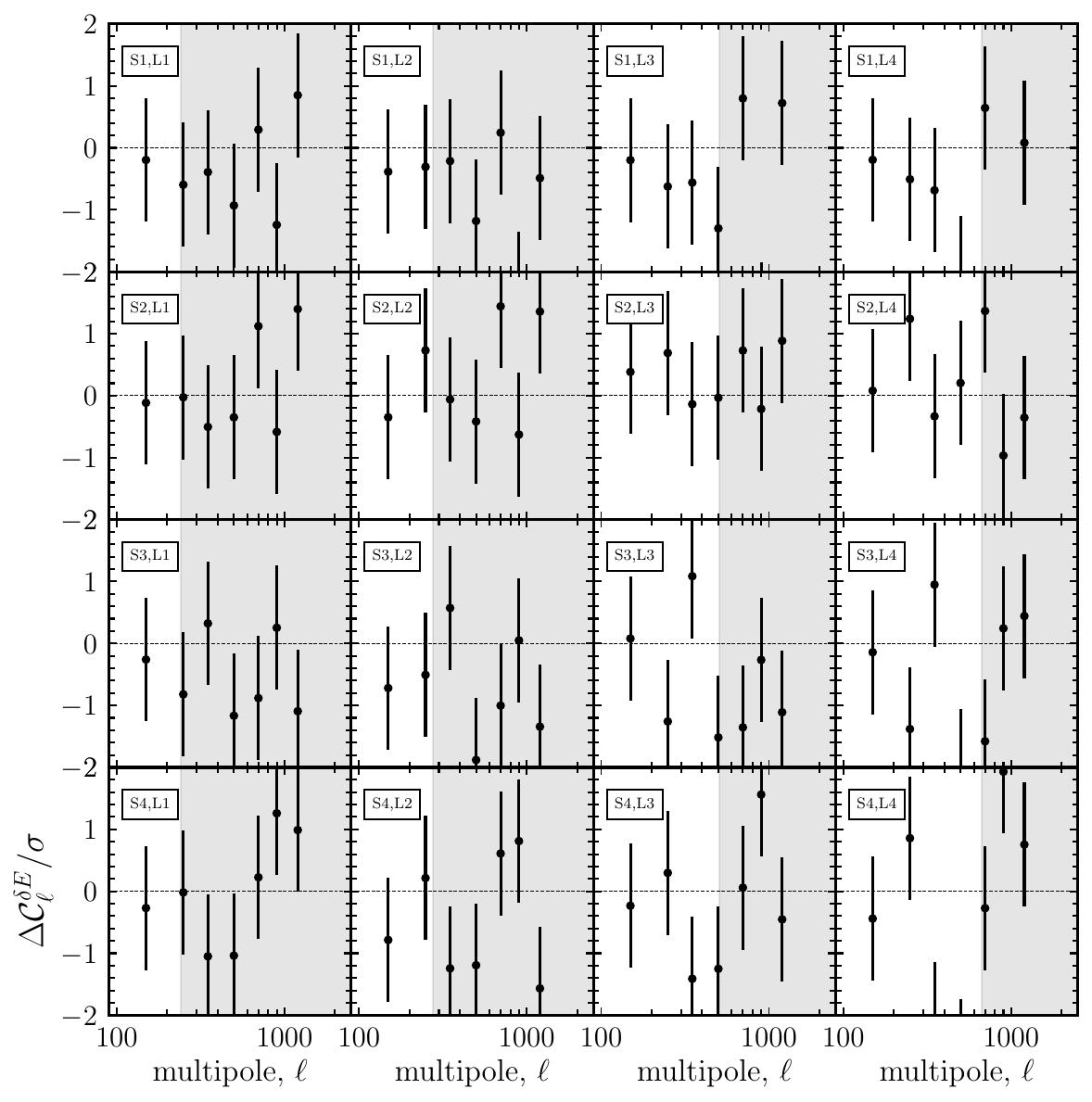}
    \caption{Impact of mode deprojection on the measured angular power spectra, expressed as the residual $\Delta C_\ell/\sigma$ between signals with and without deprojection applied, in units of the $1\sigma$ uncertainty. \textit{Left:} cosmic shear, for each tomographic bin-pair $(i,j)$. \textit{Right:} galaxy--galaxy lensing, for each source--lens bin combination Si,Lj. In both panels, deprojection effects remain within $2\sigma$ across the analyzed scales with no systematic trend toward increased or decreased power, showing no statistically significant evidence that the deprojection biases the cosmological signal. Gray shaded regions indicate scales excluded from the analysis.}
    \label{fig:deprojection}
\end{figure*}

Figure~\ref{fig:deprojection} shows the fractional difference between signals with and without mode deprojection for cosmic shear (upper panel) and galaxy--galaxy lensing (lower panel). For both probes, deprojection effects within the analyzed scale range remain within $2\sigma$ with no systematic trend toward increased or decreased power. 

The impact of mode deprojection on cosmic shear and galaxy--galaxy lensing is small (Figure~\ref{fig:deprojection}): the shear measurements are already calibrated to remove systematic effects, and only the two DCR maps are deprojected from the shear field, whose effect is expected to be minor given that Subaru is equipped with an atmospheric dispersion corrector. The absolute impact of deprojection is larger for galaxy clustering, as the density field couples to a larger set of observing-condition templates. We do not repeat the clustering analysis here and instead refer the reader to \citet{nicola2020tomographic}, who quantified this effect for the same sample: comparing clustering power spectra measured with and without deprojection (their Section~4.1.3 and Figure~7), they found differences smaller than $\sim\!0.3\sigma$ across all bin pairs, indicating that the residual contamination is small and well described by the linear deprojection model.

Additionally, in Appendix~\ref{app:deprojection-cmb} we verify that the SP maps do not correlate with the cosmological signal --- a central assumption of the deprojection method --- by cross-correlating them with \ac{CMB} lensing convergence and \ac{tSZ} maps from \textit{Planck}~\citep{Planck:2018lbu, Planck:2015vgm}. We find mean Pearson correlation coefficients of $(1.05 \pm 3.32)\times 10^{-3}$ with the $\kappa$-map and $(4.52 \pm 5.64)\times 10^{-3}$ with the tSZ map, both consistent with zero.

\section{Modeling}
\label{sec:modeling}

Accurate modeling of the $3\times2$pt signal requires accounting for theoretical systematics affecting weak lensing and galaxy clustering, as well as calibration of observational effects. On scales where theoretical modeling is incomplete, we exclude the affected data points via scale cuts to avoid biasing the cosmological constraints. Theoretical predictions are computed with \textsc{Firecrown} \textit{(in prep.)}, the \ac{DESC} inference framework, which interfaces with the \ac{CCL}~\citep{Chisari_2019} and \textsc{CAMB} \citep{Lewis_2000} for evaluation of cosmological quantities. The set of modeling choices implemented in the signal modeling pipeline is summarized in Table~\ref{tab:baseline-modeling}.

\subsection{Baseline modeling pipeline}
\label{subsec:baseline-modeling}

On small scales, non-linear gravitational evolution modifies the matter power spectrum. We estimate the non-linear dark matter-only power spectrum with \textsc{Halofit}~\citep{Takahashi_2012, Smith_2003}, a fitting formula calibrated on N-body simulations. Our primary analysis assumes massless neutrinos, following~\citet{hikage2019}. When investigating non-zero neutrino mass, we replace the standard \textsc{Halofit} power spectrum with the prescription of~\citet{Bird_2012}.

\textbf{Galaxy clustering theoretical systematics}. The main theoretical systematic affecting the clustering description on small scales is the galaxy bias \citep{Desjacques:2016bnm, LSSTDarkEnergyScience:2023qfp} --- the relation between the galaxy and matter density fields. We model this using a linear bias prescription,

\begin{equation}
    \delta_g^i = b^i_g\, \delta_{\mathrm{m}},
\end{equation}

where $b^i_g$ is the linear bias coefficient for lens redshift bin $i$, assumed constant within each bin \citep{Porredon_2022}. Scales where non-linear bias becomes significant are removed via scale cuts (Section~\ref{subsec:cuts}).

\textbf{Weak lensing theoretical systematics}. The main sources of uncertainty at small scales are baryonic feedback and \ac{IA}. Baryonic feedback --- driven by \ac{AGN} and stellar outflows --- suppresses the matter power spectrum at small scales \citep{Chisari:2019tus, Prat:2025ucy}. We mitigate this effect by applying scale cuts and adopting a dark matter-only power spectrum as our baseline. We model \ac{IA} \citep{DES:2022vuu, DESI:2025fda, Fortuna:2021apz} --- the gravitational alignment of galaxy shapes with the large-scale tidal field, which mimics the cosmic shear signal --- using the \ac{NLA} model \citep{bridle2007dark}. \ac{NLA} introduces intrinsic--intrinsic (II) and gravitational--intrinsic (GI) alignment contributions,

\begin{equation}
    P_{\mathrm{II}} = F^2(z)\,P_{m}(k,z), 
    \qquad 
    P_{\mathrm{GI}} = F(z)\,P_{m}(k,z),
\end{equation}

where $P_{m}(k,z)$ is the non-linear matter power spectrum and the redshift-dependent modulation is

\begin{equation}
    F(z) = -A_{\mathrm{IA}}\,C_1\,\rho_{\mathrm{crit},0}
    \frac{\Omega_{\mathrm{m}}}{D(z)}
    \left(\frac{1+z}{1+z_0}\right)^{\eta_{\mathrm{eff}}},
\end{equation}

with $A_{\mathrm{IA}}$ the IA amplitude, $C_1 = 5\times10^{-14} h^{-2}M_\odot^{-1}\,\mathrm{Mpc}^3$ a normalization constant, $\rho_{\mathrm{crit},0}$ the present critical density, $D(z)$ the growth factor, $\eta_{\mathrm{eff}}$ the redshift power-law index, and $z_0 = 0.62$ the pivot redshift.

\begin{table}
    \centering
    \small
    \begin{tabular}{lr}
    \toprule
    \multicolumn{2}{c}{\textbf{Modeling choices}} \\
    \midrule
    Matter power spectrum & \textsc{Halofit} DM-only $(+\sum m_\nu)$ \\
    \midrule
    \multicolumn{2}{c}{\textit{Source galaxy sample}} \\
    \midrule
    Intrinsic alignment        & NLA \\
    Shape uncertainty          & Multiplicative shear bias \\
    Photo-$z$ uncertainty      & Shift \\
    PSF leakage                & -- \\
    \midrule
    \multicolumn{2}{c}{\textit{Lens galaxy sample}} \\
    \midrule
    Galaxy bias                & Linear + scale cuts \\
    Photo-$z$ uncertainty      & Shift and stretch \\
    Lens magnification bias         & -- \\
    \bottomrule
    \end{tabular}
    \caption{Baseline modeling choices for the $3\times2$pt analysis. Extensions to massive neutrinos and magnification are explored in Section~\ref{subsec:results-extended}. Modeling ingredients not considered in the baseline pipeline are marked with ``--''.}
    \label{tab:baseline-modeling}
\end{table}

\textbf{Data calibration}. Weak lensing calibration systematics are addressed following \citet{hikage2019}. We account for multiplicative shear bias \citep{Huterer:2005ez, Heymans:2005rv} by introducing one free parameter $\Delta m$ common to all source tomographic bins, marginalized with a Gaussian prior. We additionally apply fixed corrections for multiplicative selection bias $m_{\mathrm{sel}}^i$ and responsivity $m_{\mathcal{R}}^i$ per redshift bin \citep[Table~4]{hikage2019}. These biases enter the angular power spectra as

\begin{equation}
    C^{ij}_\ell \to 
    \left(1 + \Delta m\right)^2 
    \left(1 + m^i_{\mathrm{sel}} + m^i_{\mathcal{R}}\right) 
    \left(1 + m^j_{\mathrm{sel}} + m^j_{\mathcal{R}}\right) 
    C^{ij}_\ell.
\end{equation}

We do not model \ac{PSF} leakage or residual PSF errors~\citep{Zhang_2023} owing to the absence of a publicly accessible star shape catalog measured with the same algorithm as the source~galaxies. The impact on cosmological constraints is expected to~be negligible, consistent with the findings of~\citet{hikage2019}, which showed these systematics produce insignificant changes to the~final results.

Finally, we calibrate photo-$z$ uncertainties by introducing a shift $\Delta z^{\mathrm{l/s}}$ per redshift bin for each sample,

\begin{equation}
    n^{\mathrm{l/s}}(z) = n_{\mathrm{photo}}^{\mathrm{l/s}}
    \left(z - \Delta z^{\mathrm{l/s}}\right),
\end{equation}

with Gaussian priors for the source sample and flat priors for the lenses (Table~\ref{tab:cosmological-parameters}). For the lens sample, we additionally allow a stretch of the redshift distribution \citep{krause2021dark, descollaboration2021dark},

\begin{equation}
    n^{\mathrm{l}}(z) = n_{\mathrm{photo}}^{\mathrm{l}}
    \left(\sigma_{z,l}\left[z - \langle z \rangle\right] 
    + \langle z \rangle\right),
\end{equation}

where $\sigma_{z,l}$ is the stretch parameter per lens redshift bin, with a flat prior (Table~\ref{tab:cosmological-parameters}).

\textbf{Point-mass marginalization.} In real space, the tangential shear $\gamma_t$ is non-local: measurements at angular scale $\theta$ receive contributions from all smaller scales, which needs to be accounted for in the covariance calculation~\citep{MacCrann_2019, Prat_2023} and thus have a non-negligible impact on scale cuts selection and derived cosmology. In harmonic space, the signal is decomposed into multipoles and thus has a more direct scale localization, so point-mass marginalization is not required~\citep{faga2024darkenergysurveyyear}.

\subsection{Scale cuts}
\label{subsec:cuts}

The theoretical modeling of galaxy clustering and weak lensing does not accurately describe all scales accessible to modern surveys \citep{Sugiyama_2023, miyatake2021cosmological, Wright:2025xka, DES:2026zjp}. We therefore apply scale cuts to exclude data points where the baseline assumptions are known to break down: a dark matter-only non-linear power spectrum computed with \textsc{Halofit}, neglecting baryonic feedback \citep{Amon_2022, Secco_2022, Aric__2023}; linear galaxy bias \citep{LSSTDarkEnergyScience:2023qfp, Porredon_2022, Pandey_2022}; and the \ac{NLA} model for \ac{IA}.

For cosmic shear, we analyze multipoles in the range $300 < \ell < 1900$, following \citet{hikage2019}. Scales with $\ell < 300$ show a significant excess of B-mode power \citep{hikage2019, Oguri_2017}, while the cut at $\ell > 1900$ mitigates uncertainties from \ac{NLA} modeling on small scales and baryonic feedback, permitting the use of a dark matter-only power spectrum.

\begin{figure*}[t!]
    \centering
    \includegraphics[width=0.51\textwidth]{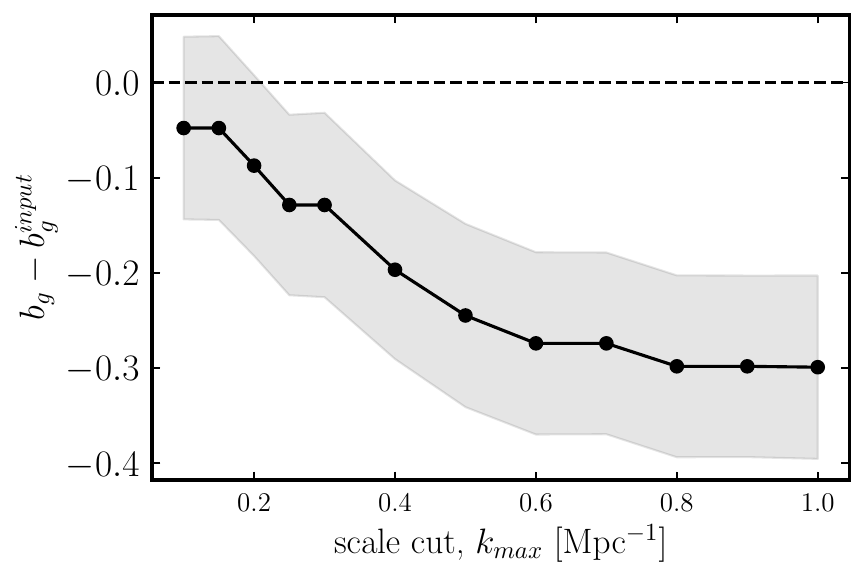}
    \includegraphics[width=0.455\textwidth]{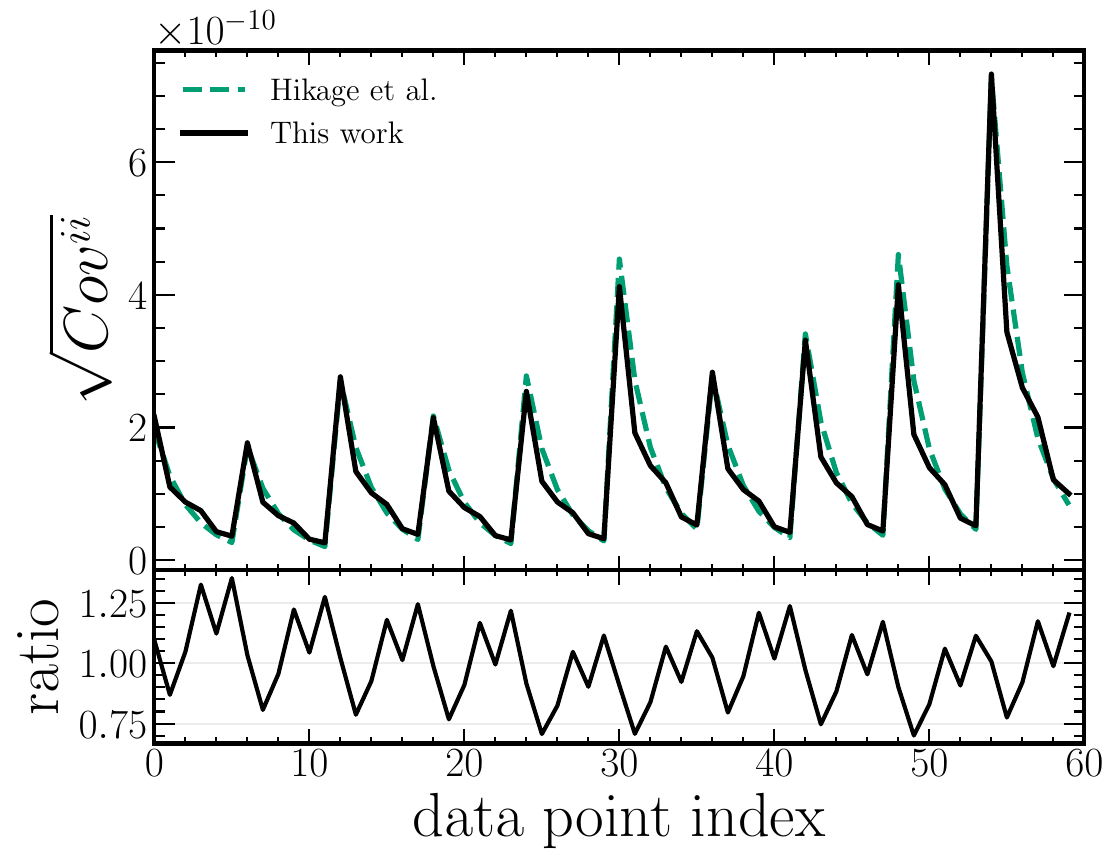}
    \caption{\textit{Left:} Difference between the estimated and input linear galaxy bias coefficient $b_g - b_g^{\rm input}$ as a function of the maximum wavenumber $k_{\rm max}$ included in the fit, for the first lens redshift bin ($z_{\rm eff} = 0.40$), where $b_g^{\rm input} = 1.17$ is the known bias of the mock HOD signal. The shaded band shows the $1\sigma$ jackknife uncertainty. The fiducial scale cut $k_{\rm max}=0.15\,{\rm Mpc}^{-1}$ is selected as the largest value (retaining the maximum number of data points) for which the recovered bias remains within the $1\sigma$ jackknife uncertainty. \textit{Right:} Comparison of the cosmic shear diagonal covariance elements $\sqrt{\mathrm{Cov}^{ii}}$ between \textsc{TJPCov} (black solid) and the covariance released by \citet{hikage2019} (green dashed). The lower panel shows the ratio of the diagonal elements, consistent to within 25\%.}
    \label{fig:clustering-scalecuts-bias}
\end{figure*}

For galaxy clustering and galaxy--galaxy lensing, we determine the smallest scale accurately described by the linear bias model via the following iterative procedure. A mock \ac{HOD} signal \citep{Berlind_2003} is generated assuming a \textit{Planck}-like cosmology \citep{aghanim2020planck} with best-fit \ac{HOD} parameters from \citet{nicola2020tomographic}. For each choice of maximum wavenumber $k_{\rm max}$, we model the mock signal with linear galaxy bias, run a minimizer to recover the bias coefficients, and compare them to the known input values $b_g = \{1.17, 1.39, 1.70, 2.30\}$ for the four lens redshift bins at $z_{\rm eff} = \{0.40, 0.67, 1.00, 1.49\}$. We repeat this for $k_{\rm max} \in \{0.1, 0.15, 0.2, 0.3, 0.5, 0.7, 1.0, 1.5\}\,{\rm Mpc}^{-1}$, and select the largest $k_{\rm max}$ (i.e.\ retaining the maximum number of data points) for which the recovered bias $b_g - b_g^{\rm input}$ remains within the 68\% jackknife credible interval. We find $k_{\rm max} = 0.15\,{\rm Mpc}^{-1}$; results for the first redshift bin are shown in Figure~\ref{fig:clustering-scalecuts-bias}.

\section{Inference Framework}
\label{section:analysis}

We adopt a Bayesian framework for parameter inference, comparing the measured data vector $\hat{\mathbf{D}}$ to the theoretical prediction

\begin{equation}
    \mathbf{T}_{\mathrm{M}}(\boldsymbol{\Theta}) \equiv 
    \left\{ C_{EE}(\boldsymbol{\Theta}, \ell),\, 
            C_{\delta_g E}(\boldsymbol{\Theta}, \ell),\, 
            C_{\delta_g \delta_g}(\boldsymbol{\Theta}, \ell) 
    \right\}
\end{equation}

for model $\mathrm{M}$ with parameters $\boldsymbol{\Theta}$. We assume a Gaussian likelihood,

\begin{equation}
    \mathcal{L}(\hat{\mathbf{D}}|\boldsymbol{\Theta}, \mathrm{M}) 
    \propto \exp\left\{-\frac{1}{2}
    \left(\hat{\mathbf{D}} - \mathbf{T}_{\mathrm{M}}\right)^\intercal 
    \mathrm{Cov}^{-1} 
    \left(\hat{\mathbf{D}} - \mathbf{T}_{\mathrm{M}}\right)
    \right\},
\end{equation}

where $\mathrm{Cov}$ is the covariance matrix of the signal described in Section~\ref{subsec:covariance}. We derive the posterior probability distribution of the model parameters as

\begin{equation}
    P(\boldsymbol{\Theta}|\hat{\mathbf{D}}, \mathrm{M}) \propto 
    \mathcal{L}(\hat{\mathbf{D}}|\boldsymbol{\Theta}, \mathrm{M})\, 
    \Pi(\boldsymbol{\Theta}|\mathrm{M}),
\end{equation}

where $\Pi(\boldsymbol{\Theta}|\mathrm{M})$ is the prior distribution over cosmological and nuisance parameters (Table~\ref{tab:cosmological-parameters}). We constrain $\Lambda$CDM and $w$CDM models, with extensions including a free neutrino mass $\sum m_\nu$. To assess goodness-of-fit, we report the reduced $\chi^2$ with an effective number of degrees of freedom

\begin{equation}
    \nu = N_{\mathrm{Data}} - N_{\mathrm{eff}},
\end{equation}

where $N_{\mathrm{Data}}$ is the number of data points and $N_{\mathrm{eff}}$ is the number of effectively constrained parameters,

\begin{equation}
    N_{\mathrm{eff}} = N_{\mathrm{params}} - 
    \mathrm{tr}\left[\mathcal{C}^{-1}_{\mathrm{prior}}\, 
    \mathcal{C}_{\mathrm{post}}\right].
    \label{eqn:neff}
\end{equation}

The cosmological inference pipeline is implemented in \textsc{Firecrown} \textit{(in prep.)} interfacing with \textsc{CosmoSIS}~\citep{Zuntz_2015} for \ac{MCMC} sampling. We employ two nested sampling configurations. To reproduce the results of \citet{hikage2019} we use \textsc{MultiNest} \citep{Feroz_2009} with \texttt{tol = 0.1}, \texttt{efr = 0.3}, and \texttt{Nlive = 200}, with constant efficiency mode disabled. For the fiducial cosmological constraints and robustness tests we use \textsc{PolyChord} \citep{Handley_2015}, configured with \texttt{Nlive = 250}, \texttt{tol = 0.01}, and \texttt{fast\_fraction = 0.0}, following the sampler optimization described in \citet{lemos2022robust}. As an indication of the computational cost of this analysis, a typical cosmic shear chain took $\sim$30 minutes on 256 CPUs and produced $\sim$3\,500 posterior samples, while a full $3\times2$pt chain took $\sim$3 hours and produced $\sim$9\,000 posterior samples. We note that the number of posterior samples in nested sampling is substantially smaller than the total number of likelihood evaluations, owing to the slice-sampling steps and rejected proposals within each iteration; we do not report the latter figure here.

\begin{table*}
    \centering
    \small
    \begin{tabular}{llcr}
    \toprule
    Model & Parameter & Symbol & Prior \\
    \midrule
    \textbf{Flat $\Lambda$CDM} 
        & Physical dark matter density & $\Omega_c h^2$                   & flat$[0.03, 0.7]$            \\
        & Physical baryon density      & $\Omega_b h^2$                   & flat$[0.019, 0.026]$         \\
        & Hubble parameter             & $h$                              & flat$[0.6, 0.9]$             \\
        & Scalar amplitude             & $\ln(10^{10} A_s)$               & flat$[1.5, 6.0]$             \\
        & Scalar spectral index        & $n_s$                            & flat$[0.87, 1.07]$           \\
    \midrule
    \textbf{$w$CDM}  
        & Dark energy EoS              & $w$                              & fixed$(-1)$ or flat$[-3, -1/3]$ \\
    \midrule
    \textbf{$\nu$CDM} 
        & Neutrino mass                & $\sum m_\nu\,[\mathrm{eV}]$      & fixed$(0)$, fixed$(0.06)$, or flat$[0, 1]$ \\
    \midrule
    \textbf{Intrinsic alignment}      
        & IA amplitude                 & $A_{\mathrm{IA}}$                & flat$[-5, 5]$                \\
        & IA redshift evolution        & $\eta_{\mathrm{eff}}$            & flat$[-5, 5]$                \\
    \midrule
    \textbf{Sources}  
        & Multiplicative shear bias    & $100\Delta m$                    & $\mathcal{N}(0, 1)$          \\
    \midrule
    \textbf{Lenses}   
        & Linear galaxy bias           & $b^{1\ldots4}_g$                 & flat$[0.1, 6.0]$             \\
        & Lens magnification bias           & $s^{1\ldots4}$                   & fixed $\{0.09, 0.29, 0.29, 0.37\}$ \\
    \midrule
    \textbf{Photo-$z$ calibration}
        & Source shift, bin $i$        & $100\Delta z^{\mathrm{s}}_i$     & $\mathcal{N}(0, \sigma_i)$, $\sigma_i = \{2.85, 1.35, 3.83, 3.76\}$ \\
        & Lens shift, bin $i$          & $\Delta z^{\mathrm{l}}_i$        & flat$[-0.5, 0.5]$, $i \in \{1,2\}$; flat$[-1.0, 1.0]$, $i \in \{3,4\}$ \\
        & Lens stretch, bin $i$        & $\sigma_{z,l}^i$                 & flat$[0.8, 1.2]$             \\
    \bottomrule
    \end{tabular}
    \caption{\label{tab:cosmological-parameters} Parameters and priors adopted for the flat $\Lambda$CDM and $w$CDM analyses. The baseline model assumes a cosmological constant ($w = -1$, fixed) and massless neutrinos ($\sum m_\nu = 0$, fixed). Extensions to $w$CDM, massive neutrinos ($\nu$CDM), and magnification are explored in Section~\ref{subsec:results-extended}; magnification bias coefficients $s^{1\ldots4}$ are fixed when included. Gaussian priors $\mathcal{N}(\mu, \sigma)$ are placed on the multiplicative shear bias and photo-$z$ shift parameters, whose widths are determined by external calibrations; all remaining parameters use uninformative flat priors.}
\end{table*}

\subsection{Covariance}
\label{subsec:covariance}
Cosmic shear, galaxy clustering, and galaxy--galaxy lensing are correlated probes, as all trace the same underlying matter density field. We compute the analytical covariance matrix of the signal~\citep{Eifler_2009, Friedrich_2021} with \textsc{TJPCov} (\textit{in prep.}), accounting for the Gaussian term \citep{cooray2001power}. We do not include the \ac{SSC} term \citep{barreira2018complete} nor the \ac{cNG} term \citep{kayo2013information}, as both contribute minimally to the error budget on the scales retained after our scale cuts \citep{barreira2018accurate}. We validate our covariance by comparing the diagonal elements of our \textsc{TJPCov} estimate with the public covariance from \citet{hikage2019}, finding agreement within 25\%, as shown in Figure~\ref{fig:clustering-scalecuts-bias}. This comparison covers only the diagonal elements of the cosmic shear block; the off-diagonal and cross-probe terms of the full 3 $\times$ 2pt covariance are not compared against an external estimate here, and their validation is presented separately in the \textsc{TJPCov} methodology paper (\textit{in prep}.). Both covariances are analytical and Gaussian-dominated --- \citet{hikage2019} show in their Figure~18 that the Gaussian term dominates the diagonal elements, confirming that the absent \ac{cNG} and \ac{SSC} terms cannot account for this difference. We instead attribute the 25\% discrepancy to methodological distinctions: their flat-sky per-field mode-coupling matrix with pixel-weighting correction $w_2^2/w_4$ versus our curved-sky \textsc{HEALPix} implementation, and their Monte Carlo shot noise estimate versus our analytical one. The ratio panel of Figure~\ref{fig:clustering-scalecuts-bias} shows no systematic trend across tomographic bin pairs; the largest deviations instead appear toward the high-multipole end of each tomographic block, where shot noise dominates and the \citet{hikage2019} covariance slightly exceeds ours, consistent with the difference between their Monte Carlo and our analytical shot-noise estimate. The impact of this covariance-level agreement on parameter precision is at a comparable order: comparing the $1\sigma$ uncertainties from our fiducial cosmic shear analysis (\textsc{TJPCov}, Eq.~\ref{eq:combined_constraints}) to those from the \textsc{Firecrown} validation using the \citet{hikage2019} published covariance (Appendix~\ref{app:firecrown-validation}, Eq.~\ref{eq:valid-tjpcov}), the $\Omega_m$ uncertainty is $\sim$31\% wider, $\sigma_8$ is essentially unchanged ($\sim$4\% narrower), and $S_8$ is $\sim$20\% narrower with \textsc{TJPCov}. These differences are consistent in scale with the $\sim$25\% covariance discrepancy itself, confirming that this level of agreement is acceptable and has negligible impact on our cosmological constraints.

\section{Results}
\label{section:results}
In this section, we present constraints on $\Lambda$CDM and $w$CDM derived from the cosmic shear, $2\times2$pt, and $3\times2$pt analyses. Alternative modeling choices such as massive neutrinos are explored to stress-test the fiducial constraints against different data configurations and calibrations. Cosmological constraints are reported as the median and 68\% credible interval of the posterior distribution.

\subsection{$\Lambda$CDM constraints}
\label{subsec:results-LCDM}

From the cosmic shear analysis of the angular power spectra measured in this work, we obtain
\begin{equation}
    \Omega_m = 0.239^{+0.111}_{-0.076}, \quad
    \sigma_8 = 0.922^{+0.186}_{-0.177}, \quad
    S_8 = 0.815^{+0.020}_{-0.023}.
    \label{eq:combined_constraints}
\end{equation}
We validate this pipeline by re-analyzing the cosmic shear signal of~\citet{hikage2019} using their published covariance matrix and our \textsc{Firecrown}-based modeling framework, finding good agreement with the publicly released HSC results; see Appendix~\ref{app:firecrown-validation} for details.

From the $2\times2$pt analysis we obtain
\begin{equation}
    \Omega_m = 0.268^{+0.062}_{-0.059}, \quad
    \sigma_8 = 0.738^{+0.132}_{-0.115}, \quad
    S_8 = 0.694^{+0.077}_{-0.071}.
\end{equation}
Constraints on the linear galaxy bias coefficients and intrinsic alignment parameters from this analysis are listed in Table~\ref{tab:hsc-lcdm-parameter-constraints} and shown in Figure~\ref{fig:results-gbias}.

\begin{figure*}
    \centering
    \includegraphics[width=0.6\linewidth]{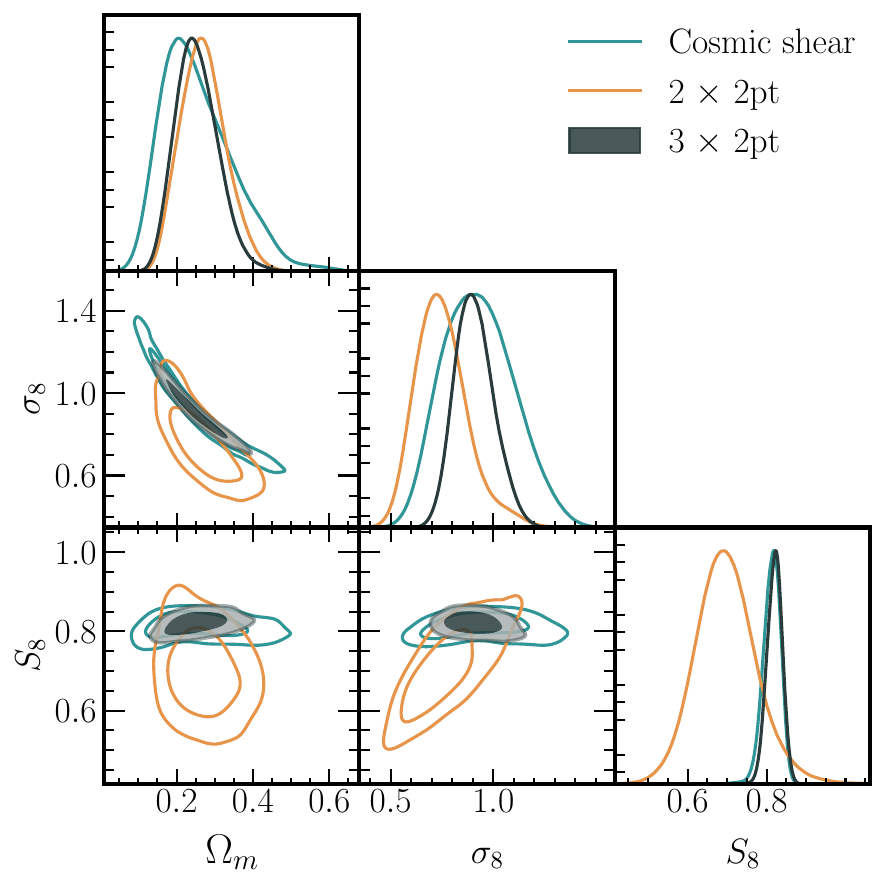}
    \caption{\label{fig:results-lcdm} Marginalized posterior distributions of $\Omega_m$, $\sigma_8$, and $S_8 \equiv \sigma_8\sqrt{\Omega_m/0.3}$ in the $\Lambda$CDM scenario, derived from the cosmic shear analysis ($1\times2$pt, teal), the joint galaxy clustering and galaxy--galaxy lensing analysis ($2\times2$pt, orange), and the full $3\times2$pt analysis (dark gray filled). Contours show the 68\% and 95\% credible levels.}
\end{figure*}

The full $3\times2$pt analysis yields
\begin{equation}
    \Omega_m = 0.249^{+0.061}_{-0.051}, \quad
    \sigma_8 = 0.904^{+0.098}_{-0.091}, \quad
    S_8 = 0.821^{+0.018}_{-0.021},
    \label{eq:3x2pt_constraints}
\end{equation}
with a reduced chi-square of $\chi^2/\nu = 105.67/111.0 = 0.95$, corresponding to a p-value of $p = 0.63$, where the effective number of constrained parameters $N_\mathrm{eff} = 3.62$. Constraints on the linear galaxy bias coefficients and photometric redshift calibration parameters for both the source and lens samples are presented in Table~\ref{tab:hsc-lcdm-parameter-constraints}. The marginalized posterior distributions of best-constrained parameters are shown in Figure~\ref{fig:results-lcdm}.

\begin{table}
    \centering
    \renewcommand{\arraystretch}{1.25}
    \resizebox{\columnwidth}{!}{%
    \begin{tabular}{lcccc}
    \toprule
    \multicolumn{5}{c}{\textbf{Cosmic shear ($1\times2$pt)}} \\
    \midrule
    All $z$-bins & \multicolumn{4}{c}{$A_{\rm IA}$ = $0.45^{+0.48}_{-0.59}$ \; \; \; \;  $\eta_{\rm eff}$ = $0.93^{+2.69}_{-3.32}$} \\[3pt]   
    \midrule\midrule
    \multicolumn{5}{c}{\textbf{$2\times2$pt}} \\
    \midrule
    Parameter & $z$-bin 1 & $z$-bin 2 & $z$-bin 3 & $z$-bin 4 \\
    \midrule
    $b_g$            & $1.26^{+0.34}_{-0.32}$ & $1.40^{+0.30}_{-0.24}$ & $1.86^{+0.38}_{-0.28}$ & $2.13^{+0.42}_{-0.35}$ \\ [3pt]   
    All $z$-bins & \multicolumn{4}{c}{$A_{\rm IA}$ = $0.91^{+0.69}_{-1.01}$ \; \; \; \;  $\eta_{\rm eff}$ = $-1.55^{+3.24}_{-2.23}$} \\ [3pt]   
    \midrule\midrule
    \multicolumn{5}{c}{\textbf{$3\times2$pt}} \\
    \midrule
    Parameter & $z$-bin 1 & $z$-bin 2 & $z$-bin 3 & $z$-bin 4 \\
    \midrule
    $b_g$
        & $1.07^{+0.29}_{-0.26}$
        & $1.14^{+0.17}_{-0.15}$
        & $1.51^{+0.20}_{-0.18}$
        & $1.74^{+0.23}_{-0.21}$ \\ [3pt]
    $100\,\Delta z^{\mathrm{s}}$
        & $-1.2^{+2.6}_{-2.4}$
        & $-0.1^{+1.2}_{-1.3}$
        & $0.7^{+2.9}_{-2.8}$
        & $2.3^{+3.2}_{-3.3}$ \\ [3pt]
    $100\,\Delta z^{\mathrm{l}}$
        & $3.6^{+7.7}_{-7.2}$
        & $1.4^{+5.8}_{-5.2}$
        & $20.1^{+7.7}_{-5.9}$
        & $21.6^{+13.7}_{-11.2}$ \\ [3pt]
    $\sigma_{z}^{\mathrm{l}}$
        & $0.987^{+0.136}_{-0.129}$
        & $1.048^{+0.103}_{-0.141}$
        & $0.998^{+0.135}_{-0.135}$
        & $1.021^{+0.124}_{-0.136}$ \\ [3pt]
    All $z$-bins & \multicolumn{4}{c}{$A_{\rm IA}$ = $0.32^{+0.43}_{-0.48}$ \; \; \; \;  $\eta_{\rm eff}$ = $-0.65^{+3.27}_{-2.79}$} \\ [3pt]
    \bottomrule
    \end{tabular}
    }
    \caption{\label{tab:hsc-lcdm-parameter-constraints} Marginalized $\Lambda$CDM constraints (68\% credible interval) on nuisance parameters from the cosmic shear ($1\times2$pt), $2\times2$pt, and $3\times2$pt analyses. Per-bin parameters ($b_g$, $\Delta z^{\rm s}$, $\Delta z^{\rm l}$, $\sigma_z^{\rm l}$) are listed for each of the four lens or source tomographic redshift bins; photo-$z$ shift parameters $\Delta z$ are multiplied by 100 for readability. Global parameters ($A_{\rm IA}$, $\eta_{\rm eff}$) are shared across all $z$-bins.}
\end{table}

\subsection{Extended models}
\label{subsec:results-extended}

\begin{figure*}
    \centering
    \includegraphics[width=0.9\textwidth]{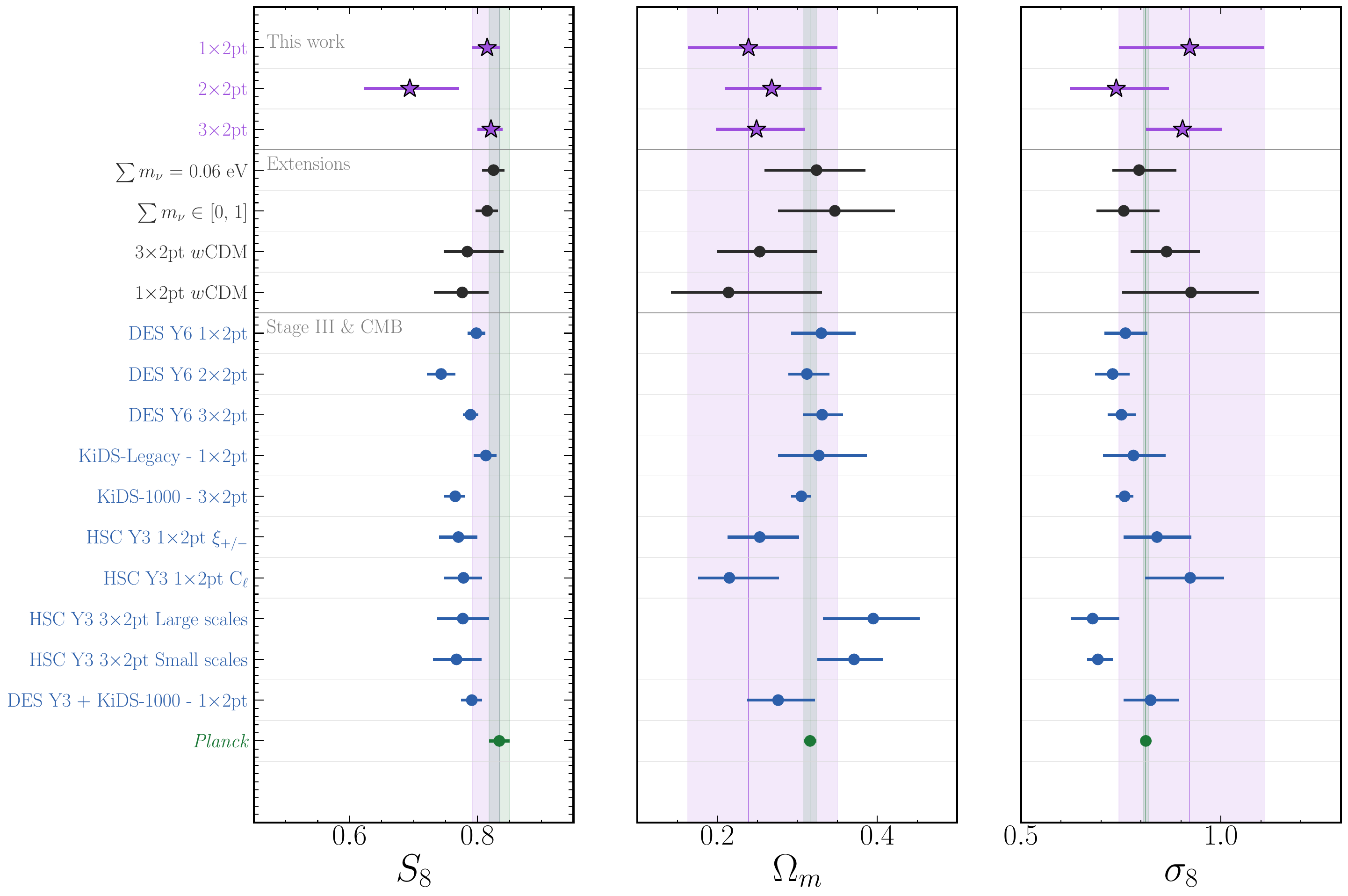}
    \caption{Marginalized constraints on $S_8$, $\Omega_m$, and $\sigma_8$ from this work and from the literature, reported as the median and 68\% credible interval. Results are grouped into three blocks: ``This work'' (purple) shows the fiducial $1\times2$pt, $2\times2$pt, and $3\times2$pt constraints; ``Extensions'' (black) shows modeling variants including massive neutrinos and $w$CDM; ``Stage~III \& CMB'' (blue and green) shows constraints from \ac{DES} Year~6~\citep{DES:2026mkc,DES:2026fyc}, KiDS-1000~\citep{Heymans:2020gsg} and Legacy~\citep{Wright:2025xka}, HSC Year~3~\citep{Sugiyama_2023, miyatake2023hyper}, and the \textit{Planck} 2018 \ac{CMB} temperature and polarization analysis~\citep{aghanim2020planck}. A direct comparison of constraining power with Stage~III surveys is not straightforward, as this analysis was conducted in an unblinded manner with modeling choices adjusted to replicate the HSC~Y1 cosmic shear results.}
    \label{fig:hsc-results-literature}
\end{figure*}

\begin{figure}[!tb]
    \centering
    \includegraphics[width=0.9\linewidth]{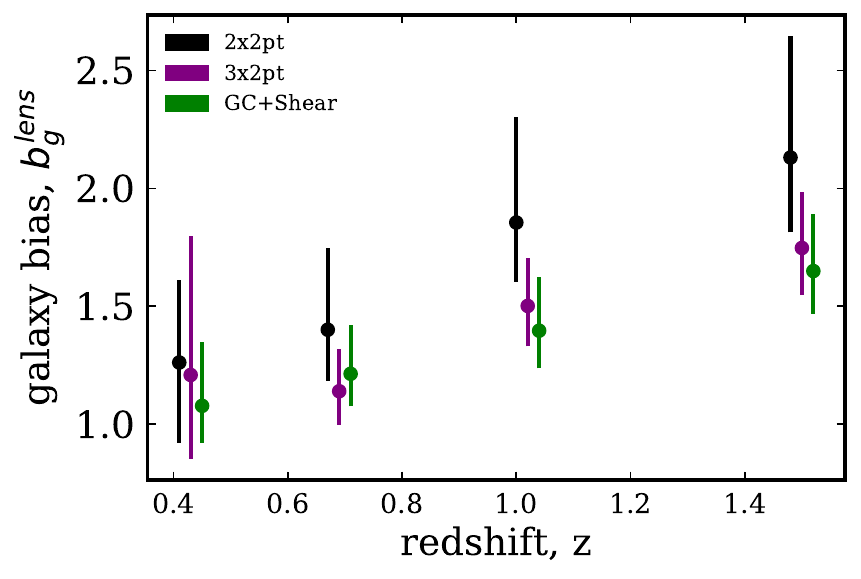}
    \caption{\label{fig:results-gbias} Linear galaxy bias coefficients $b_g^{\rm lens}$ constrained at the four lens tomographic bin effective redshifts $z_{\rm eff} \in \{0.40, 0.67, 1.00, 1.49\}$, from the $2\times2$pt (black), $3\times2$pt (purple), and GC$+$Shear (green) analyses. Error bars show the $1\sigma$ uncertainty. All three methods yield consistent bias values, with the $3\times2$pt and GC$+$Shear analyses generally producing tighter constraints than the $2\times2$pt alone.}
\end{figure}

We explore three extensions beyond the fiducial modeling choices; results are summarized in the ``Extensions'' panel of Figure~\ref{fig:hsc-results-literature}.

We first allow the dark energy equation-of-state parameter $w$ to vary freely in $[-3, -1/3]$. Both the cosmic shear and $3\times2$pt analyses show a tendency toward lower $\sigma_8$ and $S_8$, though all results remain consistent with the fiducial $\Lambda$CDM values within $1\sigma$. We find no significant evidence for a departure from a cosmological constant.

We also consider massive neutrinos under two scenarios: masses fixed at $\sum_\nu m_\nu = 0.06\,\mathrm{eV}$, and masses free to vary in $[0.0,\,1.0]\,\mathrm{eV}$. Both are consistent with the fiducial massless-neutrino model. We observe a mild shift toward lower $\Omega_{\rm m}$ and higher $\sigma_8$, while the median $S_8$ remains well within $1\sigma$ of the fiducial result.

Finally, we assess the contribution of lens magnification to the  observed galaxy density contrast. We introduce a magnification bias term given by Eq.~\ref{eq:magnif-kernel}, with per-bin coefficients $s = \{0.09,\, 0.29,\, 0.29,\, 0.37\}$ estimated in \citet{nicola2020tomographic} from the slope of the observed cumulative $i$-band magnitude distribution of the \ac{HSC}~Y1 Wide lens sample at $m_{\rm lim} = 24.5$, following Eq.~\ref{eq:magnif-slope}. Including in the model the lens magnification contribution to the galaxy overdensity, with coefficients fixed to the values from \citet{nicola2020tomographic}, yields $\chi^2 = 104.41$ compared to $\chi^2 = 105.67$ for the baseline model. As the magnification coefficients are fixed rather than sampled, this is a comparison of the goodness-of-fit between two models with the same number of free parameters. The resulting change, $\Delta\chi^2 = 1.26$, is much smaller than the expected root-mean-square fluctuation of the $\chi^2$ statistic for $\nu \approx 111$ degrees of freedom, $\sqrt{2\nu}\approx 14.9$, indicating that the assumed magnification amplitude has a negligible impact on the goodness-of-fit of our baseline analysis.

\subsection{Robustness tests}
\label{subsec:results-robustness}

\begin{figure*}
    \centering
    \includegraphics[width=0.9\textwidth]{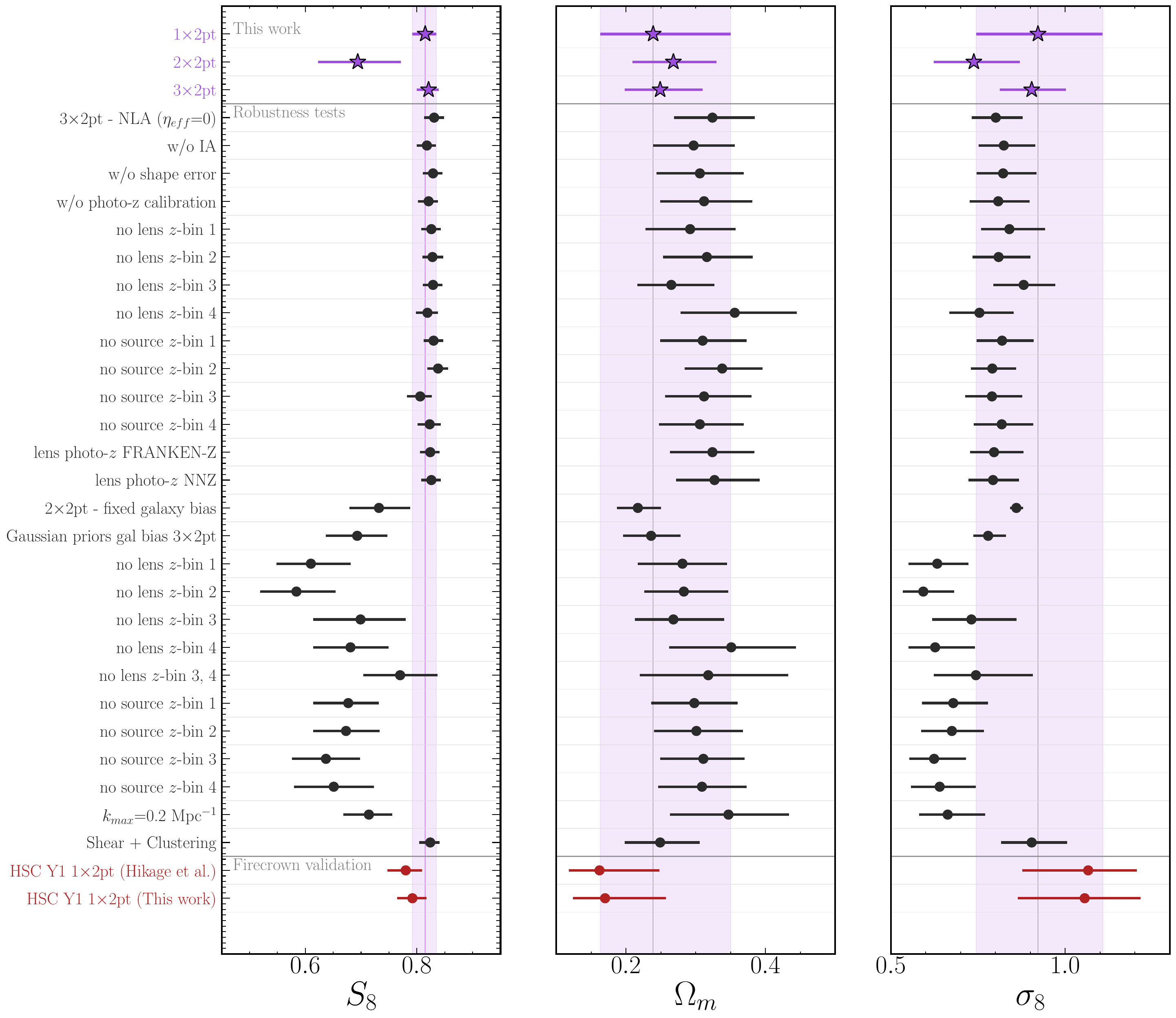}
    \caption{\label{fig:hsc-robustness-tests} Summary of $\Lambda$CDM constraints on $S_8$, $\Omega_m$, and $\sigma_8$, reported as the median and 68\% credible interval. Results are grouped into three blocks. ``This work'' (purple, top three rows): fiducial $1\times2$pt, $2\times2$pt, and $3\times2$pt constraints, detailed in Section~\ref{subsec:results-LCDM}. ``Robustness tests'' (black): sensitivity tests of the $3\times2$pt analysis (rows from ``$3\times2$pt -- NLA with $\eta_{\rm eff} = 0$'' through ``Shear + Clustering'') and of the $2\times2$pt analysis (rows from ``$2\times2$pt -- fixed galaxy bias'' onward), covering IA modeling, calibration of galaxy shape error via multiplicative shear bias, photo-$z$ calibration, tomographic bin selection, scale cuts, and galaxy bias assumptions; only the row labelled ``Gaussian priors gal bias $3\times2$pt'' imposes Gaussian priors on the bias parameters, while all remaining rows use the fiducial flat priors. ``Firecrown validation'' (red): comparison between the official HSC~Y1 cosmic shear result \citep{hikage2019} and our re-analysis of their published data vector with \textsc{Firecrown}, validating the likelihood implementation. The vertical band shows the $1\sigma$ region of the fiducial $1\times2$pt constraint on each parameter.}
\end{figure*}

We stress-test our baseline results against different data configurations and modeling choices to identify potential inconsistencies in the theoretical pipeline and dataset. Results are shown in the ``Robustness Tests'' panel of Figure~\ref{fig:hsc-robustness-tests}.

The first set of tests concerns the stability of the $3\times2$pt results. Removing the redshift dependence of the NLA intrinsic alignment model, or eliminating the \acs{IA} signal entirely, has negligible impact on the constraints, consistent with the limited sensitivity of our dataset to intrinsic alignments. Results are similarly stable when excluding combinations of lens and source tomographic bins, fixing the photo-$z$ shift and shape error parameters, or adopting alternative methods for estimating the lens redshift distributions. The full set of configuration variants is shown in Figure~\ref{fig:hsc-robustness-tests}. Across all robustness tests, the fiducial $3\times2$pt constraints remain stable within $1\sigma$.

We also examine the consistency of the $2\times2$pt constraints. Excluding any combination of source tomographic bins leaves the results consistent with the fiducial analysis, while removing higher-redshift lens bins produces a trend toward lower $S_8$ values. We also test different galaxy bias assumptions: fixing the bias values, or imposing Gaussian priors centered on the $3\times2$pt posterior medians with widths set by the posterior uncertainties. Constraints from both tests are consistent with the fiducial 2$\times$2pt case, with smaller error bars thanks to the additional information in the priors. We note that this test is exploratory, and the resulting constraints should be interpreted with caution: since the priors are derived from the 3$\times$2pt posterior, which itself includes the 2$\times$2pt data, this configuration is not an independent cross-check of the galaxy bias, and the improved precision reflects this circularity rather than new external information. We further test the sensitivity to scale cuts by relaxing the galaxy clustering and galaxy-galaxy lensing cuts to $k_{\rm max} = 0.2\,\mathrm{Mpc}^{-1}$, which introduces a bias toward lower $S_8$ values, confirming that the fiducial scale cuts are necessary to avoid nonlinear contamination.

As an additional configuration, we construct an alternative $2\times2$pt data vector combining cosmic shear with galaxy clustering instead of galaxy-galaxy lensing. This combination is dominated by the cosmic shear constraining power and yields results closer to the cosmic shear and $3\times2$pt analyses. It also provides a third independent constraint on the galaxy bias parameters, complementing those from the standard $2\times2$pt and $3\times2$pt analyses (see Figure~\ref{fig:results-gbias}).

\section{Conclusions}
\label{section:conclusions}

We have presented the first fully photometric $3\times2$pt analysis of \ac{HSC} Y1 data in harmonic space, performed using \ac{LSST}-\ac{DESC} analysis pipelines as a direct precursor to the forthcoming \ac{LSST}-\ac{DESC} cosmological analysis. Using photometrically selected source and lens galaxy samples processed independently of the official HSC pipeline, we measured the angular power spectra of cosmic shear, galaxy clustering, and galaxy--galaxy lensing across a $136.9\,\mathrm{deg}^2$ partial-sky footprint, and combined them into a joint cosmological inference.

A key methodological contribution of this work is the application of mode deprojection with \textsc{TXPipe} to remove observational systematics encoded in spin-2 \ac{SP} maps that directly affect the shear field. We confirmed that the deprojection does not remove cosmological signal by cross-correlating \ac{SP} maps with CMB maps (Appendix~\ref{app:deprojection-cmb}), finding residual effects well within $2\sigma$. Null tests --- the BB, EB, and BE modes for cosmic shear and the B-mode power spectrum for galaxy--galaxy lensing --- show no statistically significant evidence for B-mode contamination. The covariance matrix, computed with \textsc{TJPCov} accounting for the Gaussian term, was used for cosmological inference carried out with \textsc{Firecrown}.

From the $3\times2$pt analysis we obtain $\Omega_m = 0.249^{+0.061}_{-0.051}$ and $S_8 = 0.821^{+0.018}_{-0.021}$, with a goodness-of-fit $\chi^2/\nu = 105.67/111.0 = 0.95$ ($p = 0.63$). The robustness of these constraints was tested against a broad range of modeling and data configuration variants, including different tomographic bin combinations, IA model choices, scale cut thresholds, and redshift distribution estimation methods, with all results remaining consistent within $1\sigma$. We validated \textsc{Firecrown} by reproducing the HSC~Y1 cosmic shear results of \citet{hikage2019} (Appendix~\ref{app:firecrown-validation}).

We also explored extensions beyond $\Lambda$CDM, allowing the dark energy equation-of-state parameter $w$ to vary, freeing or fixing the neutrino mass sum, and including lens magnification. None of these extensions shows significant evidence for departure from the baseline model.

Our $3\times2$pt constraints are consistent with those from the leading Stage-III weak-lensing surveys, as summarized in Figure~\ref{fig:hsc-results-literature}. The DES Year~6 cosmic shear \citep{DES:2026mkc}, $2\times2$pt and $3\times2$pt \citep{DES:2026fyc} analyses in real space are all compatible with our findings; these analyses apply scale cuts to limit baryonic feedback contamination and adopt a matter power spectrum with a fixed baryonic strength and linear and non-linear galaxy bias~\citep{DES:2026zjp}. Our results are similarly consistent with the \ac{KiDS}-1000 $3\times2$pt \citep{Heymans:2020gsg} and KiDS-Legacy cosmic shear~\citep{Wright:2025xka} constraints, with KiDS preferring a marginally lower $S_8$ while statistically consistent. From HSC Year~3, the cosmic shear analyses in real \citep{dalal2023hyper} and harmonic space \citep{li2023hyper}, and the large- \citep{Sugiyama_2023} and small-scale \citep{miyatake2023hyper} $3\times2$pt analyses with SDSS spectroscopic lenses, are all in good agreement with our results. We also find consistent constraints with the DES Year~3 $\times$ KiDS-1000 joint cosmic shear re-analysis \citep{2023}. We note that a direct comparison of constraining power is not straightforward, as this analysis was conducted in an unblinded manner with modeling choices adjusted to replicate the HSC Year~1 cosmic shear results. Our constraints are further compared against the \textit{Planck} CMB primary anisotropy results \citep{aghanim2020planck} in Figure~\ref{fig:hsc-results-literature}.

This work demonstrates the capability of LSST-DESC pipelines for end-to-end joint analyses of weak lensing and galaxy clustering, and establishes the infrastructure foundation for the forthcoming LSST-DESC $3\times2$pt analysis. The tools and techniques validated here are directly deployable on Stage-III legacy datasets and on the first generations of LSST data.

\section*{Acknowledgments}
\label{sec:ackownledgment}

DSC acknowledges support from the Ministry of Science and Education of Spain through research grant PRE2019-088032, the University of Zurich (UZH) and the Swiss National Science Foundation (SNSF) under grant number 10002981. We thank the LSST Corporation for financial support through the 2021-02 Enabling Science Program, which enabled DSC to visit JS at Fermilab. We also thank researchers at the Kavli Institute for Cosmological Physics (KICP) at the University of Chicago for valuable discussions. CGG was supported by the Beecroft Trust and the C\'esar Nombela  Research Talent Attraction grant from the Community of Madrid  (Ref. 2025-T1/TEC-36302). N\v{S} is supported in part by the OpenUniverse effort, which is funded by NASA under JPL Contract Task 70-711320, ‘Maximizing Science Exploitation of Simulated Cosmological Survey Data Across Surveys’. This research used resources of the National Energy Research Scientific Computing Center (NERSC), supported by the Office of Science of the U.S. Department of Energy under Contract No. DE-AC02-05CH11231. The DESC acknowledges ongoing support from the Institut National de Physique Nucl\'eaire et de Physique des Particules in France; the Science \& Technology Facilities Council in the United Kingdom; and the Department of Energy and the LSST Discovery Alliance in the United States.  DESC uses resources of the IN2P3 Computing Center (CC-IN2P3--Lyon/Villeurbanne - France) funded by the Centre National de la Recherche Scientifique; the National Energy Research Scientific Computing Center, a DOE Office of Science User Facility supported by the Office of Science of the U.S.\ Department of Energy under Contract No.\ DE-AC02-05CH11231; STFC DiRAC HPC Facilities, funded by UK BEIS National E-infrastructure capital grants; and the UK particle physics grid, supported by the GridPP Collaboration. This work was performed in part under DOE Contract DE-AC02-76SF00515. We thank Roohi Dalal for comments and for sharing the HSC Year~3 cosmic shear posterior samples. This paper has undergone internal review in the LSST Dark Energy Science Collaboration. We are grateful to the internal reviewers, Boris Leistedt and Masaya Yamamoto, who provided helpful feedback and suggestions on the manuscript.

\section*{Author contributions}
\label{sec:author-cont}

D. Sanchez-Cid, J. Sanchez: Project co-leadership; formal analysis;
writing of the manuscript.
I. Sevilla-Noarbe, D. Alonso, F. Andrade-Oliveira, H. Awan, C. Chang,
J. Ellison, C. Garc\'ia-Garc\'ia, E. Longley-Phillips, R. Mandelbaum,
A. Nicola, J. Prat, E. Sanchez, M. Soares-Santos, J. Zuntz: Overall guidance and feedback.
F. Andrade-Oliveira, C. Garc\'ia-Garc\'ia: \textsc{TJPCov}
development and covariance methodology guidance.
E. Rykoff: \textsc{Decasu} \ac{SP} map production.
M. Yamamoto: Internal review.
E. Pedersen: \textsc{TXPipe} developer and DESC Builder.
M. Ishak, N. Sarcevic: DESC Builders.

\section*{Software and Data Availability}
\label{sec:software}

All measurements and analyses were performed on the Perlmutter
platform at the National Energy Research Scientific Computing Center
(NERSC), utilizing the DESC computing and storage allocation.

This work made use of the following software:
\begin{itemize}
    \item \textsc{TXPipe} \citep{prat2022catalog}: DESC end-to-end
    pipeline for two-point lensing and clustering statistics in real
    and harmonic space (\url{https://github.com/LSSTDESC/TXPipe}).
    \item \textsc{TJPCov} (\textit{in prep.}): analytical covariance matrix calculator (\url{https://github.com/LSSTDESC/tjpcov}).
    \item \textsc{Firecrown}: DESC likelihood framework (\url{https://github.com/LSSTDESC/firecrown}) interfacing
    with \textsc{CosmoSIS} (\citealt{Zuntz_2015}, \url{https://cosmosis.readthedocs.io/en/latest/}), \textsc{CCL}
    (\citealt{Chisari_2019}, \url{https://github.com/LSSTDESC/CCL}), and \textsc{CAMB} \citep{Lewis_2000}.
    \item \textsc{NaMaster} \citep{Alonso_2019}: pseudo-$C_\ell$
    power spectrum estimator with mode deprojection (\url{https://github.com/LSSTDESC/NaMaster}).
    \item \textsc{Sacc}: the DESC standard format for storing
    redshift distributions, measurements, and covariances in a
    unified object, used to store the $3\times2$pt data vector
    and covariance matrix produced in this work (\url{https://github.com/LSSTDESC/sacc}).
    \item \textsc{Decasu}: \ac{SP} map generator (\url{https://github.com/erykoff/decasu}).
\end{itemize}

The data products derived from this analysis, including angular power
spectra, analytical covariance matrix, redshift distributions, and
\textsc{CosmoSIS} chain outputs for the cosmic shear, $2\times2$pt,
and $3\times2$pt analyses, will be shared upon reasonable request to
the corresponding author.

\section*{Affiliations}

\begin{list}{}{\leftmargin=1.2em \itemindent=-1.2em \itemsep=0pt \parsep=0pt}
\item $^{1}$ Centro de Investigaciones Energ\'eticas, Medioambientales y Tecnol\'ogicas (CIEMAT), Madrid, Spain
\item $^{2}$ Physik-Institut, University of Z\"urich, Winterthurerstrasse 190, CH-8057 Z\"urich, Switzerland
\item $^{3}$ Space Telescope Science Institute, 3700 San Martin Dr, Baltimore, MD 21218, USA
\item $^{4}$ Department of Physics, University of Oxford, Denys Wilkinson Building, Keble Road, Oxford OX1 3RH, United Kingdom
\item $^{5}$ Kavli Institute for Particle Astrophysics and Cosmology, Department of Physics, Stanford University, Stanford, CA 94309, USA
\item $^{6}$ SLAC National Accelerator Laboratory, 2575 Sand Hill Road, Menlo Park, CA 94025, USA
\item $^{7}$ Department of Astronomy and Astrophysics, University of Chicago, Chicago, IL 60637, USA
\item $^{8}$ Kavli Institute for Cosmological Physics, University of Chicago, Chicago, IL 60637, USA
\item $^{9}$ NSF-Simons AI Institute for the Sky (SkAI), 172 E. Chestnut St., Chicago, IL 60611, USA
\item $^{10}$ University of California Riverside, Riverside, CA 92521, USA
\item $^{11}$ Waterloo Centre for Astrophysics, University of Waterloo, Waterloo, ON N2L 3G1, Canada
\item $^{12}$ Department of Physics and Astronomy, University of Waterloo, Waterloo, ON N2L 3G1, Canada
\item $^{13}$ Department of Physics, The University of Texas at Dallas, Richardson, Texas 75080, USA
\item $^{14}$ Physics Department, Duke University, Durham, NC 27708, USA
\item $^{15}$ McWilliams Center for Cosmology and Astrophysics, Department of Physics, Carnegie Mellon University, Pittsburgh, PA 15213, USA
\item $^{16}$ Jodrell Bank Centre for Astrophysics, Department of Physics and Astronomy, The University of Manchester, Manchester M13 9PL, United Kingdom
\item $^{17}$ School of Mathematics, Statistics and Physics, Newcastle University, Herschel Building, NE1 7RU Newcastle-upon-Tyne, U.K.
\item $^{18}$ Nordita, KTH Royal Institute of Technology and Stockholm University, Hannes Alfv\'ens v\"ag 12, SE-10691 Stockholm, Sweden
\item $^{19}$ University of Copenhagen, Dark Cosmology Centre, Juliane Maries Vej 30, 2100 Copenhagen O, Denmark
\item $^{20}$ Princeton University, Department of Astrophysical Sciences, 4 Ivy Ln, Princeton, NJ, 08544, USA
\item $^{21}$ Institute for Astronomy, University of Edinburgh, Blackford Hill, Edinburgh EH9 3HJ, United Kingdom
\end{list}

\bibliographystyle{mnras}
\bibliography{bibliography_re}

\newpage
\null
\clearpage
\appendix

\section{Source and lens galaxy samples}
\label{app:galaxy-numbers}

Table~\ref{tab:hsc-data-number-galaxies} lists the number of source and lens galaxies in each of the five \ac{HSC} fields and tomographic redshift bins used in this analysis. Source galaxies span $0.3 < z < 1.5$ across four tomographic bins, and lens galaxies span $0.15 < z < 1.5$, totaling \num{8813360} source and \num{8235621} lens galaxies over $136.9$ sq. deg.

\begin{table}[h]
    \centering
    \resizebox{\columnwidth}{!}{
    \begin{tabular}{lcccc}
    \toprule
    \multicolumn{5}{c}{\textbf{Source galaxies}} \\
    \midrule
    Field & $0.3 < z < 0.6$ & $0.6 < z < 0.9$ & $0.9 < z < 1.2$ & $1.2 < z < 1.5$ \\
    \midrule
    \textsc{Gama09h} & \num{754785}  & \num{700547}  & \num{523454}  & \num{300463} \\
    \textsc{Gama15h} & \num{692816}  & \num{700349}  & \num{519927}  & \num{303458} \\
    \textsc{Vvds}    & \num{424470}  & \num{441723}  & \num{334718}  & \num{194616} \\
    \textsc{Wide12h} & \num{292930}  & \num{299044}  & \num{228273}  & \num{134503} \\
    \textsc{Xmm}     & \num{593494}  & \num{617977}  & \num{480856}  & \num{274957} \\
    \midrule
    Total            & \num{2758495} & \num{2759640} & \num{2087228} & \num{1207997} \\
    \midrule\midrule
    \multicolumn{5}{c}{\textbf{Lens galaxies}} \\
    \midrule
    Field & $0.15 < z < 0.5$ & $0.5 < z < 0.75$ & $0.75 < z < 1.0$ & $1.0 < z < 1.5$ \\
    \midrule
    \textsc{Gama09h} & \num{959257}  & \num{871308}  & \num{822952}  & \num{865774} \\
    \textsc{Gama15h} & \num{721598}  & \num{736852}  & \num{686957}  & \num{706758} \\
    \textsc{Vvds}    & \num{363880}  & \num{355340}  & \num{370092}  & \num{379156} \\
    \textsc{Wide12h} & \num{246889}  & \num{253816}  & \num{248898}  & \num{262074} \\
    \textsc{Xmm}     & \num{520055}  & \num{515383}  & \num{512644}  & \num{543383} \\
    \midrule
    Total            & \num{2024826} & \num{2090874} & \num{2014036} & \num{2105885} \\
    \bottomrule
    \end{tabular}
    }
    \caption{\label{tab:hsc-data-number-galaxies} Number of source and lens galaxies per \ac{HSC} field and tomographic redshift bin.}
\end{table}

\section{Cross-correlation of survey property maps with CMB tracers}
\label{app:deprojection-cmb}

A key assumption of the mode deprojection method is that the \ac{SP} maps encoding observational systematics are not intrinsically correlated with the cosmological signal. We verify this by cross-correlating our \ac{SP} maps with two \textit{Planck} tracers of the large-scale structure \citep{aghanim2020planck}: the CMB lensing convergence ($\kappa$-map,~\citet{Carron:2022eyg}), and the thermal Sunyaev--Zel'dovich (\acs{tSZ},~\citet{Planck:2015vgm}) Compton $y$-map. A non-zero correlation would indicate that the deprojection is removing cosmological signal alongside the systematics. A similar cross-correlation test between systematic maps and external \ac{LSS} tracers is performed in \citet{Weaverdyck2026} in the context of the \ac{DES} Year~6 analysis (their Figure~14).

\textbf{CMB lensing} --- The lensing convergence~\citep{Hanson:2009kr} traces the projected matter distribution along the line of sight,
\begin{equation}
    \kappa(\theta) = \frac{3H_0^2\Omega_{\rm m}}{2c^2}
    \int_0^{\chi_{\rm max}} \delta(\chi,\theta)\,
    \frac{(\chi_{\rm max} - \chi)\,\chi}{\chi_{\rm max}}\,
    \mathrm{d}\chi,
\end{equation}

where $\delta(\chi,\theta)$ is the matter density field and $\chi_{\rm max}$ is the maximum comoving distance considered \citep{kovacs2022dark}. Cross-correlating the $\kappa$-map with each \ac{SP} map yields a mean Pearson correlation coefficient of $(1.05 \pm 3.32)\times 10^{-3}$, consistent with zero.

\textbf{Thermal Sunyaev--Zel'dovich} --- The Compton $y$-parameter encodes inverse Compton scattering of \ac{CMB} photons by hot intracluster electrons,

\begin{equation}
    y(\mathbf{n}) = \int \frac{k_B T_e}{m_e c^2}\,
    n_e\,\sigma_T\,\mathrm{d}s,
\end{equation}

where $n_e$, $m_e$, and $T_e$ are the electron number density, mass, and temperature, and $\sigma_T$ is the Thomson cross-section. The \ac{tSZ}~\citep{Birkinshaw:1998qp} signal traces massive halos and provides a complementary probe of the large-scale structure. Cross-correlating the $y$-map with each SP map yields a mean Pearson correlation coefficient of $(4.52 \pm 5.64)\times 10^{-3}$, also consistent with zero.

\begin{figure}[h]
    \centering
    \includegraphics[width=0.8\linewidth]{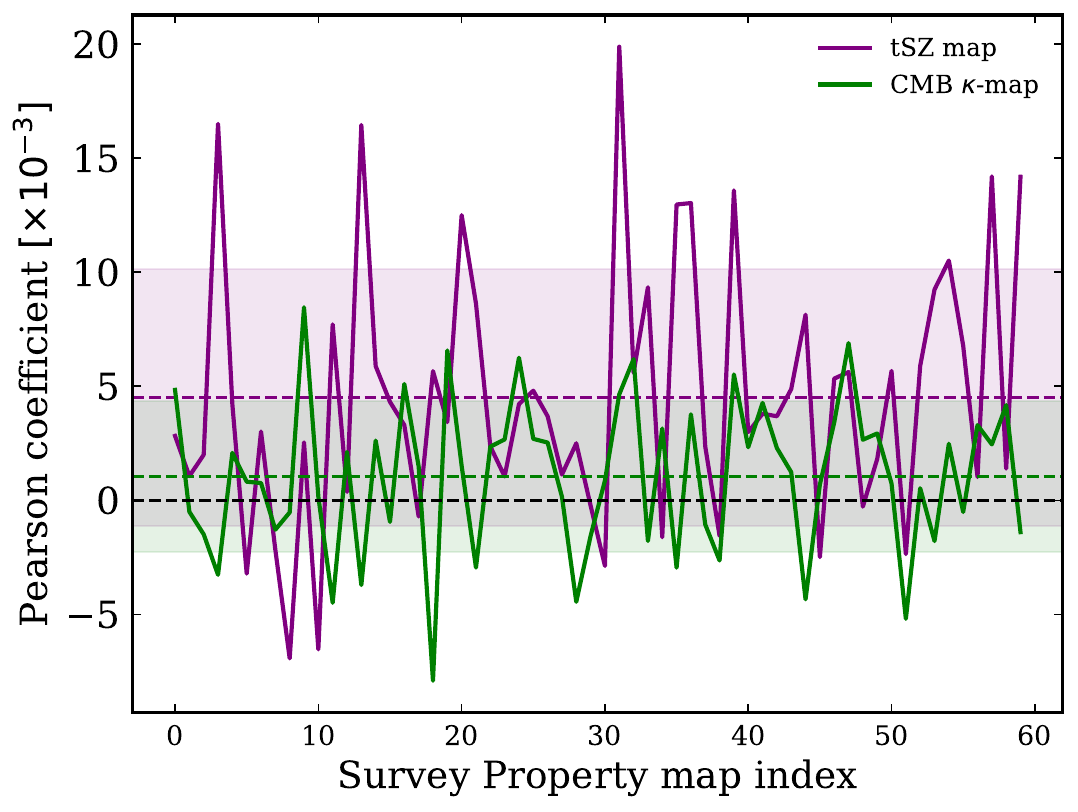}
    \caption{\label{fig:correlation_cmb_sp_maps} Pearson correlation coefficients between each \textsc{Decasu} \ac{SP} map and two \textit{Planck} tracers of large-scale structure: the CMB lensing convergence $\kappa$-map (green) and the \acs{tSZ} Compton $y$-map (purple). Dashed lines indicate the mean correlation coefficient for each tracer; shaded bands show the $1\sigma$ uncertainty. All coefficients are consistent with zero, confirming that the SP maps are not intrinsically correlated with the cosmological signal.}
\end{figure}

The individual SP map correlations for both probes are shown in Figure~\ref{fig:correlation_cmb_sp_maps}. The results show no statistically significant correlation between the \ac{SP} maps and the cosmological signal, supporting the core assumption of our deprojection procedure.

\section{Reproducing official HSC Y1 cosmic shear results}
\label{app:firecrown-validation}

\begin{figure*}
    \centering
    \includegraphics[width=0.9\linewidth]{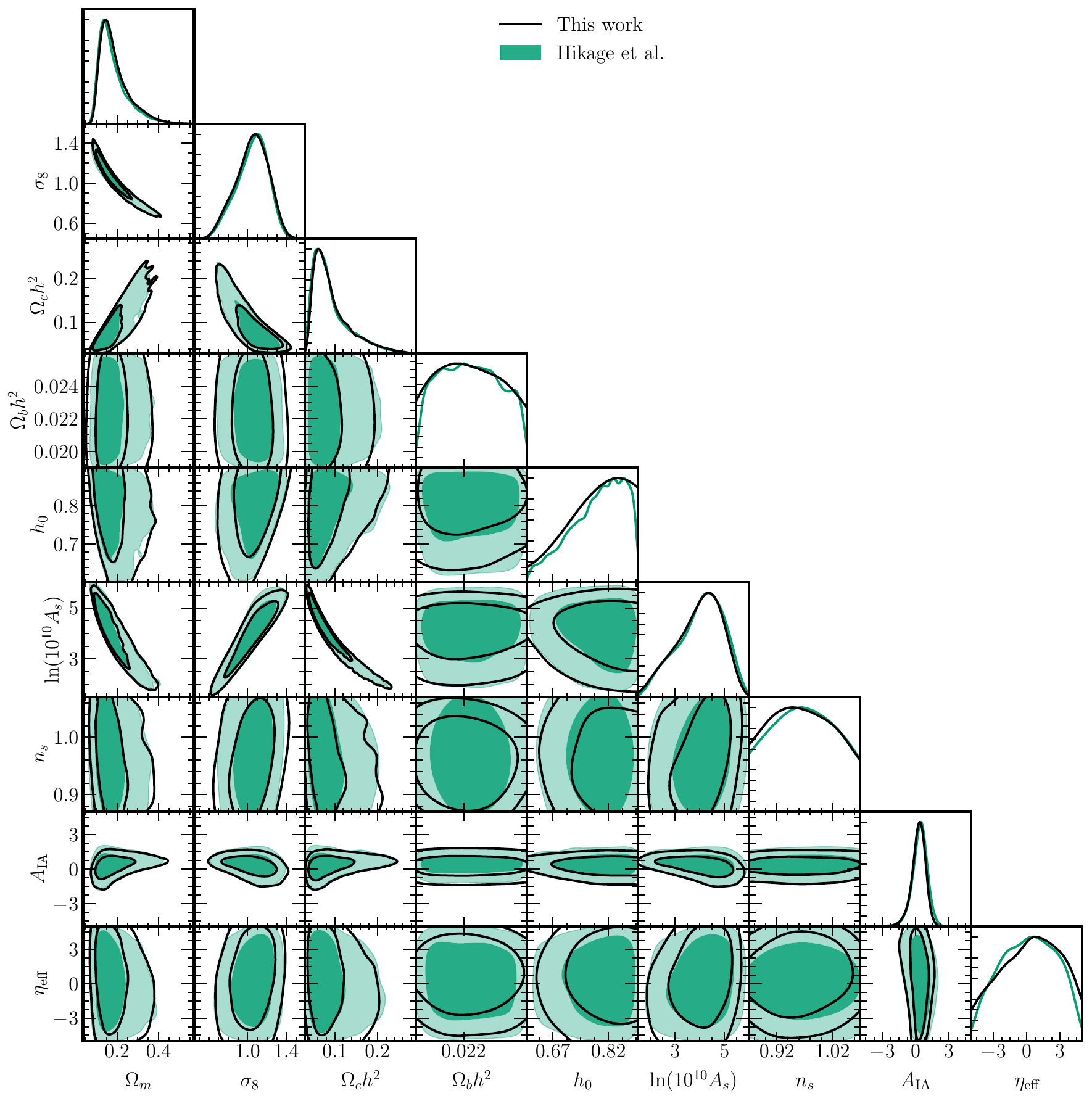}
    \caption{\label{fig:hsc-full-hikage} Validation of \textsc{Firecrown} against the official HSC Year~1 cosmic shear analysis in harmonic space \citep{hikage2019}. Contours show the 68\% and 95\% credible levels for \citet{hikage2019} (teal) and our re-analysis of their published data vector using \textsc{Firecrown} (black). Sampled parameters shown are $\Omega_b h^2$, $\Omega_c h^2$, $h$, $\ln(10^{10}A_s)$, and $n_s$; derived parameters are $\Omega_m$, $\sigma_8$, $A_{\rm IA}$, and $\eta_{\rm eff}$. The two sets of contours are in good agreement within sampling noise, validating the \textsc{Firecrown} likelihood implementation.}
\end{figure*}

We validate \textsc{Firecrown} by re-analyzing the official HSC Y1 cosmic shear angular power spectra and covariance matrix released by the HSC Collaboration\footnote{\url{https://hsc.mtk.nao.ac.jp/ssp/data-release/}} \citep{hikage2019}, following their modeling choices. The resulting posterior distributions are compared to the official results in Figure~\ref{fig:hsc-full-hikage}, showing good agreement within sampling noise.

Our re-analysis yields
\begin{equation}
    \Omega_m = 0.165^{+0.100}_{-0.043},\;
    \sigma_8 = 1.067^{+0.155}_{-0.224},\;
    S_8 = 0.789^{+0.025}_{-0.029},
    \label{eq:valid-tjpcov}
\end{equation}
in good agreement with the values reported by \citet{hikage2019},
\begin{equation}
    \Omega_m = 0.162^{+0.086}_{-0.044},\;
    \sigma_8 = 1.066^{+0.140}_{-0.189},\;
    S_8 = 0.780^{+0.029}_{-0.033}.
\end{equation}

\end{document}